\documentclass[preprint,aps,pra,showpacs,floatfix]{revtex4-1}
\usepackage[utf8]{inputenc}
\usepackage[OT1]{fontenc}
\usepackage{amsmath}
\usepackage{amsfonts}
\usepackage{amssymb}
\usepackage{bm}
\usepackage{color}
\usepackage{graphicx, subfigure}
\usepackage[left=2cm,right=2cm,top=2cm,bottom=2cm]{geometry}
\usepackage{booktabs}
\usepackage{longtable}
\usepackage{siunitx}

\allowdisplaybreaks

\newcommand{\veps}{\varepsilon}
\newcommand{\balpha}{\bm{\alpha}}

\newcommand{\br}{\bm{r}}
\newcommand{\bx}{\bm{x}}

\newcommand{\bp}{\bm{p}}

\newcommand{\be}{\begin{eqnarray}}
\newcommand{\ee}{\end{eqnarray}}

\newcommand{\psum}{\sideset{}{'}\sum}
\newcommand{\dvec}[2]{{\left( \begin{array}{c} #1 \\ #2 \\ \end{array} \right)}}

\newcommand{\non}{\nonumber \\}

\newcommand{\ket}[1]{|#1\rangle}

\newcommand{\matrixel}[3]{\big\langle#1\big|\vphantom{#1}#2\vphantom{#3}\big|#3\big\rangle}

\usepackage{color}
\definecolor{BLUE}{rgb}{0.0,0.0,1.0}

\usepackage{soul}
\soulregister\cite7
\soulregister\ref7

\begin{document}

\title{QED calculation of electron-electron correlation effects \\ in heliumlike ions}


\author{Y. S. Kozhedub}

\author{A. V. Malyshev}
\email[Corresponding author: ]{a.v.malyshev@spbu.ru}

\author{D. A. Glazov}

\author{V. M. Shabaev}

\author{I. I. Tupitsyn}

\affiliation{Department of Physics, St.~Petersburg State University, Universitetskaya 7/9, 199034 St.~Petersburg, Russia}

\begin{abstract}
Fully relativistic approach to evaluate the correlation effects in highly charged ions is presented. The interelectronic-interaction contributions of first and second orders in $1/Z$ are treated rigorously within the framework of bound-state quantum electrodynamics, whereas the calculations of the third- and higher-order contributions are based on the Dirac-Coulomb-Breit Hamiltonian. The developed approach allows one to deal with single as well as degenerate or quasi-degenerate states. We apply this approach to the calculations of the correlation contributions to the $n=1$ and $n=2$ energy levels in heliumlike ions. 
The obtained contributions are combined with the one-electron and screened QED corrections, nuclear recoil and nuclear polarization corrections to get the total theoretical predictions for the ionization and transition energies in high-$Z$ heliumlike ions.
\end{abstract}

\maketitle

\section{Introduction \label{sec:0}}
Heavy highly charged ions provide a unique opportunity for testing bound-state quantum electrodynamics (QED) in the strong-field regime. Nowadays, the accuracy of the Lamb shift measurements in H-like uranium has achieved a level of few percent of the total QED contribution~\cite{Stoehlker:2000:3109,Gumberidze:2005:223001}. Even better precision is obtained in experiments aiming to probe the QED effects in Li-like uranium \cite{Schweppe:1991:1434,Beiersdorfer:1998:3022,Bosselmann:1999:1874,Brandau:2003:073202,Beiersdorfer:2005:233003}. In order to meet the constantly improving accuracy in existing \cite{Beiersdorfer:1998:1944, Draganic:2003:183001, Gumberidze:2004:203004, Trassinelli:2007:129, Trassinelli:2009:63001, Trassinelli:2011:014003, Mackel:2011:143002, Bernhardt:2015:144008, Kraft-Bermuth:2017:055603} and planned \cite{Beyer:2015:144010, Hengstler:2015:014054, Lestinsky:2016:797, Repp:2012:983,Roux:2012:997} experiments with highly charged ions, a number of rigorous QED calculations have been performed (see, e.g., Refs.~\cite{Yerokhin:2015:033103, Yerokhin:2001:032109, Kozhedub:2007:012511, Kozhedub:2010:042513,Sapirstein:2011:012504, Malyshev:2014:062517, Malyshev:2015:012514, Artemyev:2007:173004, Artemyev:2013:032518, Malyshev:2017:022512} and references therein). Whenever possible, theoretical predictions have been compared with the results of high-precision measurements, and good agreement has been found.

Heliumlike ions play a special role among other highly charged ions. 
Possessing only two bound electrons, they represent the simplest system where the many-electron QED effects can be studied. The calculations of the ground and $n=2$ singly excited energy levels in He-like ions performed in Ref.~\cite{Artemyev:2005:062104} more than ten years ago are considered as a benchmark theoretical treatment of these effects in two-electron systems. In Ref.~\cite{Artemyev:2005:062104}, all two-electron QED corrections up to the second order of the perturbation theory were evaluated within the rigorous QED approach without an expansion in the parameter $\alpha Z$ ($\alpha$ is the fine structure constant, $Z$ is the nuclear charge number). A review of the previous relativistic calculations of heliumlike ions and a comparison with the experimental data available at that time can be found in Ref.~\cite{Artemyev:2005:062104}. Later, drawing on the x-ray transition measurements in He-like titanium and statistical treatment of the previous experimental data, Chantler \textit{et al.}~\cite{Chantler:2012:153001, Chantler:2014:123037} claimed that a divergence between the experimental results and the theory from Ref.~\cite{Artemyev:2005:062104} growing as $Z^3$ takes place. New measurements of the transition energies in middle-$Z$ heliumlike ions have been undertaken~\cite{Rudolph:2013:103002, Schlesser:2013:022503, Kubicek:2014:032508, Beiersdorfer:2015:032514, Epp:2015:020502_R, Machado:2018:032517}, and the obtained results fall outside the $Z^3$-trend predicted in Refs.~\cite{Chantler:2012:153001, Chantler:2014:123037}. Moreover, new statistical studies \cite{Epp:2013:159301, Beiersdorfer:2015:032514, Machado:2018:032517}, which include the extended sets of experimental data, have shown that there is no evidence for the aforementioned $Z$-dependent deviation. Finally, in our recent study \cite{Malyshev:2019:010501_R} we have performed completely independent \textit{ab initio} calculations of the x-ray transitions in heliumlike argon, titanium, iron, copper, and krypton. We found no possible explanation from the theoretical side for the significant discrepancy between the theory and measurements with heliumlike ${\rm Ti}^{20+}$ performed in Ref.~\cite{Chantler:2012:153001}. On the other hand, 
our results were found to be generally in agreement with the most recent high-precision experimental values. 

In the present study, we are focused on the description of the method used in Ref.~\cite{Malyshev:2019:010501_R} in order to evaluate the contribution of the correlation effects to the binding energies of He-like ions. Study of the correlation effects in heliumlike ions has a long history. There are many relativistic electronic-structure calculations performed within the lowest-order relativistic (Breit) approximation using the Dirac-Coulomb-Breit Hamiltonian~\cite{Drake:1988:586, Johnson:1992:R2197, Chen:1993:3692, Plante:1994:3519, Cheng:1994:247, Indelicato:1995:1132, Cheng:2000:044503, Watanabe:2005:074322}. The most advanced QED treatment of the correlation effects includes the two-photon exchange contribution~\cite{Blundell:1993:2615, Lindgren:1995:1167, Mohr:2000:052501, Andreev:2001:042513, Asen:2002:032516, Andreev:2003:012503, Andreev:2004:062505, Artemyev:2005:062104}. In the present work, the numerical approach employed in Ref.~\cite{Artemyev:2005:062104} has been revised thoroughly, and the introduced modifications are discussed below. We perform the rigorous evaluation of the correlation effects for $n=1$ and $n=2$ energy levels in several heliumlike ions. The contributions of the first and second orders in $1/Z$ are taken into account to all orders in $\alpha Z$. The higher-order corrections are treated within the Breit approximation using the large-scale configuration interaction (CI) method and the recursive perturbation theory (PT). 
The developed method is suitable for both single and (quasi-)degenerate levels. In comparison with Ref.~\cite{Malyshev:2019:010501_R}, the calculations are extended to  high-$Z$ ions including heliumlike uranium. An important feature of the present study is the systematic estimation of uncertainties of the obtained results. The evaluated interelectronic-interaction contributions to the binding energies are compared with the previous calculations. Our results are in agreement with those, but have much higher accuracy.
In addition, the calculations of the electron-electron correlation effects are supplemented with the evaluation of the one-electron and screened QED corrections as well as nuclear recoil and nuclear polarization contributions. This allows us to extend \textit{ab initio} QED calculations of the $n=1$ and $n=2$ energy levels performed in Ref.~\cite{Malyshev:2019:010501_R} to high-$Z$ region. The results obtained for the ionization and transition energies are compared with the previous evaluation by Artemyev \textit{et al.}~\cite{Artemyev:2005:062104} and the experimental data~\cite{Trassinelli:2009:63001, Trassinelli:2011:014003}.

The paper is organized as follows. In Section~\ref{sec:1:1}, we describe our \textit{ab initio} QED approach to evaluate the first- and second-order interelectronic-interaction effects in highly charged ions. Section~\ref{sec:1:2} is devoted to the description of two independent methods (CI and PT) to treat the higher-order correlation contributions within the Breit approximation. In Section~\ref{sec:2:1}, the numerical results for the electron-electron interaction contributions to the $n=1$ and $n=2$ energy levels in He-like ions are presented, and the comparison with the previous theoretical calculations is given. In Section~\ref{sec:2:2}, the QED calculations of the ionization and transition energies in high-$Z$ heliumlike ions are performed. The relativistic units ($\hbar=1$, $c=1$) and the Heaviside charge unit ($\alpha=e^2/4\pi$, $e<0$) are used throughout the paper.

\section{Methods of calculations \label{sec:1}}

\subsection{QED formalism \label{sec:1:1}}

The two-time Green's function (TTGF) method \cite{TTGF} represents a convenient approach to construct the QED perturbation theory for energy levels in highly charged ions. The natural zeroth-order approximation for the corresponding perturbation series is provided by the Furry picture~\cite{Furry:1951:115} with the unperturbed  many-electron relativistic wave functions defined in the $jj$~coupling. In case of two-electron ions, these functions read as
\be
\label{eq:2el_wf}
|u_i\rangle \, = A_N \sum_{m_{i_1}m_{i_2}} \langle \, j_{i_1} m_{i_1} \, j_{i_2} m_{i_2} \, | JM_J \rangle \,
\sum_{P} (-1)^P | Pi_1 Pi_2 \rangle \, ,
\ee
where $A_N$ is the normalization factor equal to $1/\sqrt{2}$ for non-equivalent electrons and $1/2$ for equivalent electrons, $j_{i_1}$ and $j_{i_2}$ are the one-electron angular momenta, $m_{i_1}$ and $m_{i_2}$ are their projections, $J$ is the total angular moment, $M_J$ is its projection, 
$\langle \, j_{i_1} m_{i_1} \, j_{i_2} m_{i_2} \, | JM_J \rangle \,$ are the Clebsch-Gordan coefficients,
$P$ is the permutation operator
$$
\sum_P (-1)^P \, | Pi_1 Pi_2 \rangle = | i_1 i_2 \rangle - | i_2 i_1 \rangle \, ,
$$
and $|i_1i_2\rangle$ is the product of one-electron wave functions $\varphi_{i_1}(\bx_1)$ and $\varphi_{i_2}(\bx_2)$ obtained from the Dirac equation,
\begin{align}
\label{eq:dirac}
h^{\rm D} \varphi_i \equiv
\big[ 
\balpha\cdot\bp + \beta m + V_{\rm nucl}
\big] 
\varphi_i
= \veps_i 
\varphi_i  \, .
\end{align}
In Eq.~(\ref{eq:dirac}), $\balpha$ and $\beta$ are the Dirac matrices, $\bp$ is the momentum operator, and $V_{\rm nucl}$ is the potential of the nucleus. Therefore, in the Furry picture the electron-nucleus interaction is taken into account to all orders in $\alpha Z$ from the very beginning. The interaction with the quantized electromagnetic field and the interelectronic interaction are considered by the perturbation theory in the parameters $\alpha$ and $1/Z$, respectively~\cite{Mohr:1998:227, TTGF, Lindgren:2004:161, Andreev:2008:135, Glazov:2011:71, Shabaev:2018:60}. We note that for very heavy ions the parameters $\alpha$ and $1/Z$ become comparable in magnitude, therefore, all the contributions can be classified by the powers of $\alpha$.

The TTGF method allows one to derive the formal expressions for the QED corrections for both single and (quasi-)degenerate states. In case of a single level, for each QED effect the TTGF method assigns some contribution which has to be included into the total binding energy of the considered state additively. This approach works well for the single states such as
$(1s\,1s)_0$, $(1s\,2s)_0$, $(1s\,2s)_1$, $(1s\,2p_{1/2})_0$, and $(1s\,2p_{3/2})_2$. 
However, evaluating the energies for the $n=2$ states in heliumlike ions, along with the single levels listed above, one encounters also the quasi-degenerate levels 
$(1s\,2p_{1/2})_1$ and $(1s\,2p_{3/2})_1$
which are split only by the relativistic effects. The energies of a set of $s$~(quasi-)degenerate states can be determined by diagonalizing the $s \times s$ matrix $H$ (in the case under consideration $s=2$). This matrix plays the role of the Hamiltonian acting in the subspace of the unperturbed (quasi-)degenerate states. It is constructed by the perturbation theory in $\alpha$ and $1/Z$ and has to include all the relevant contributions. Note, that a single level can be considered as a particular case of the set of degenerate levels with~$s=1$.

Let us formulate briefly the basic ideas how to construct within the TTGF method the matrix~$H$ for a set of $s$ (quasi-)degenerate levels with unperturbed energies $E_1^{(0)},\ldots,E_s^{(0)}$. As usual, we assume that the energy shifts of the levels under consideration are much smaller than the distance to other levels. For generality, we consider a $N$-electron ion, while for heliumlike ions $N=2$. The detailed description of the method can be found, e.g., in Refs.~\cite{TTGF, Shabaev:1993:4703, Shabaev:1994:4521}. 
The fundamental object of the method is the two-time Green's function defined as
\begin{align}
\label{eq:TTGF:G}
G( t',t ; \bx'_1, \ldots , \bx'_N ; \bx_1, \ldots , \bx_N ) 
= 
\langle 0 | T \psi(x'_1) \cdots \psi(x'_N) \bar{\psi}(x_N) \cdots \bar{\psi}(x_1) | 0 \rangle 
\left
|_{\substack{ x_1^{\prime 0}=\ldots=x_N^{\prime 0}\equiv t^\prime \\
              x_1^{0}=\ldots=x_N^{0}\equiv t }}
\right.
\, ,
\end{align}
where  $\psi$ is the electron-positron field operator in the Heisenberg representation, $\bar{\psi} = \psi^\dagger\gamma^0$, $x=(x^0,\bx)$, and $T$~is the time-ordering operator. The perturbation theory for the two-time Green's function~$G$ is formulated by means of the transition to the interaction picture. For the subsequent derivation, it is convenient to define the Fourier transform of the two-time Green's function~(\ref{eq:TTGF:G}) by
\begin{align}
\label{eq:G_TTGF}
\mathcal{G} (E; \bx'_1, \ldots , \bx'_N ; &\, \bx_1, \ldots , \bx_N ) \delta (E-E')   \nonumber \\
&=
\frac{1}{2\pi i} \frac{1}{N!}
\int_{-\infty}^\infty \! dt dt^\prime \,\, 
e^{iE^\prime t^\prime -iEt} \,
G( t',t ; \bx'_1, \ldots , \bx'_N ; \bx_1, \ldots , \bx_N ) \, .
\end{align}

The unperturbed wave functions $\left\{u_j\right\}_{j=1}^s$ of the (quasi-)degenerate levels form the $s$-dimensional subspace $\Omega$. Denoting the projector on $\Omega$ by
\be
\label{eq:quasi_P0}
P^{(0)} = \sum_{j=1}^s |u_j \rangle \langle u_j | \, ,
\ee
one can introduce the projection of the Green's function~(\ref{eq:G_TTGF}) on $\Omega$ as follows
\be 
\label{eq:P0_G_P0}
g(E) = P^{(0)} \mathcal{G} (E) \gamma_1^0 \ldots \gamma_N^0 P^{(0)} \, ,
\ee
where the spatial coordinates are omitted for brevity and the integration is implicit. Employing the Green's function~(\ref{eq:P0_G_P0}) one can determine the operators $\hat{K}$ and $\hat{P}$ by the expressions
\be
\label{eq:TTGF:K}
\hat{K} &\equiv& \frac{1}{2\pi i} \oint_\Gamma \! dE \,\, E g(E) \, , \\[2mm]
\label{eq:TTGF:P}
\hat{P} &\equiv& \frac{1}{2\pi i} \oint_\Gamma \! dE \,\,   g(E) \, .
\ee
The anticlockwise oriented  contour~$\Gamma$ in the complex $E$ plane surrounds  all $s$ levels under consideration and keeps outside all other singularities of $g(E)$. It can be shown (see Ref.~\cite{TTGF} for the detailed derivation) that the system of the (quasi-)degenerate levels is described by the operator~$\hat{H}$ defined as
\be
\label{eq:TTGF:H}
\hat{H} = \hat{P}^{-1/2} \, \hat{K} \, \hat{P}^{-1/2} \, . 
\ee
The perturbation theory for the Green's function~(\ref{eq:TTGF:G}) leads to the perturbation series for the operator~$\hat{H}$. The exact energies $E_1,\ldots,E_s$ of the states arising from the (quasi-)degenerate levels with energies $E_1^{(0)},\ldots,E_s^{(0)}$ can be found from the equation
\be
\label{eq:detH}
\det(E-H) = 0 \, ,
\ee
where $H$ is the $s \times s$ matrix with elements determined by $H_{ik} = \langle u_i | \hat{H} | u_k \rangle$.

\begin{figure}
\begin{center}
\includegraphics[height=3.5cm]{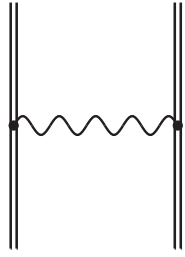}
\caption{\label{fig:1ph}
The diagram of the one-photon exchange.}
\end{center}
\end{figure}
\begin{figure}
\begin{center}
\includegraphics[width=6.3cm]{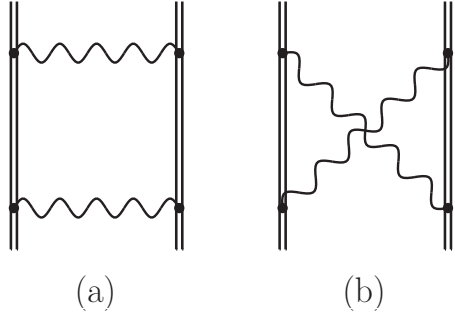}
\caption{\label{fig:2ph_2el}
The two-electron diagrams of the two-photon exchange.}
\end{center}
\end{figure}
 
To date, state-of-the-art QED calculations of the energy levels in highly charged ions comprise all contributions up to the second order in $\alpha$ and $1/Z$. The present study is devoted to the evaluation of the correlation effects. For convenience, we collect here the final expressions for the first- and second-order contributions due to the interelectronic-interaction effects which can be derived within the TTGF method. The corresponding Feynman diagrams for heliumlike ions are depicted in Figs.~\ref{fig:1ph} and~\ref{fig:2ph_2el}, respectively. A double line represents the electron propagator in the field of the nucleus, while a wavy line corresponds to the photon propagator. These diagrams do not contain any self-energy or vacuum-polarization loop and arise naturally in the QED as well as non-QED approaches (the contribution of the diagram in Fig.~\ref{fig:2ph_2el}(b) vanishes identically in the latter case). We present the formulas for the quasi-degenerate states only, but their reduction to the single-level case is straightforward.

In the $jj$-coupling scheme, the $n=2$ quasi-degenerate energy levels in He-like ions can be written as $|u_1\rangle =|\,(1s\,2p_{1/2})_1\,\rangle$ and $|u_2\rangle=|\,(1s\,2p_{3/2})_1\,\rangle$. In what follows, the indices $i$ and $k$ enumerate these states (we remind that now $s=2$ and $i,k=1,2$). In particular, the unperturbed energies $E_i^{(0)}$ are given by the sum of the one-electron Dirac energies (\ref{eq:dirac}),
\be
\label{eq:E1E2_quasi}
E_1^{(0)} = \veps_{1s} + \veps_{2p_{1/2}}, \qquad E_2^{(0)} = \veps_{1s} + \veps_{2p_{3/2}} \, .
\ee
Within the zeroth-order approximation, the Green's function (\ref{eq:P0_G_P0}) is 
\begin{align}
\label{eq:g0}
g^{(0)}(E)= \sum_{j=1}^s \frac{|u_j\rangle\langle u_j|}{E-E_{j}^{(0)}} \, .
\end{align}
Therefore, one readily obtains for the zeroth-order contribution to the $2 \times 2$ matrix $H$,   
\be
\label{eq:H0_ik}
H_{ik}^{(0)} = E_i^{(0)} \delta_{ik}  \, .
\ee
The derivation of the correction corresponding to the one-photon exchange diagram in Fig.~\ref{fig:1ph} also does not pose any difficulties. The contribution of the interelectronic-interaction effects of first order in $1/Z$ to the matrix $H$ can be represented by the expression
\be 
\label{eq:H1_ik}
H_{ik}^{(1)} = F_i F_k \, \frac{1}{2} \sum_{P} (-1)^P 
   \left[ I_{Pi_1Pi_2k_1k_2} (\Delta_1) + I_{Pi_1Pi_2k_1k_2} (\Delta_2) \right] \, , 
\ee
where $I_{abcd}(\omega) = \langle ab| I(\omega)|cd\rangle$, $I(\omega)=e^2 \alpha^\mu_1 \alpha^\nu_2 D_{\mu\nu}(\omega)$, $\alpha^\nu=\gamma^0\gamma^\mu=(1,\balpha)$, $D_{\mu\nu}(\omega)$ denotes the photon propagator, $\Delta_1 = \veps_{Pi_1}-\veps_{k_1}$ and $\Delta_2 = \veps_{Pi_2}-\veps_{k_2}$, and, for brevity, we use the following notation
\be
\label{eq:F_clebsch_def}
F_i \, |Pi_1Pi_2\rangle \equiv 
\sum_{m_{i_1}m_{i_2}}
\langle \, j_{i_1} m_{i_1} \, j_{i_2} m_{i_2} \, | JM_J \rangle \, |Pi_1Pi_2\rangle  \, 
\ee
for the summation over the angular momentum projections with the Clebsch-Gordan coefficients.

For the quasi-degenerate levels, the derivation of the formal expressions for the two-photon exchange diagrams depicted in Fig.~\ref{fig:2ph_2el} is a very complicated problem.
As compared to the single-level case, the rigorous formulas obtained within the TTGF method for the off-diagonal elements of the matrix $H$ contain 
additional terms with double integration over the energy parameters instead of the standard single integration for the diagonal elements and the single levels. This makes the direct evaluation of the second-order QED corrections for the quasi-degenerate states rather difficult and time-consuming. However, in Ref.~\cite{TTGF} it was noticed that these extra terms contribute at the level of the higher-order QED corrections. Our consideration of the QED effects is restricted to the first and second orders of the perturbation theory. Therefore, these terms are neglected within the present calculations. 

Further simplification of the calculation formulas for the off-diagonal matrix elements which was suggested in Ref.~\cite{Artemyev:2005:062104} is associated with the replacement of the zeroth-order energies $E_i^{(0)}$ and $E_k^{(0)}$ with their average value $\bar{E}_{ik}^{(0)} = (E_i^{(0)}+E_k^{(0)})/2$. For diagonal $i=k$ matrix elements, this transformation is identical and does not change the expressions. Similar to the extra integrals, a small variation of the off-diagonal elements introduced by this replacement can be disregarded as belonging to the higher-order QED effects, see also the discussion below. It is essential that the replacement introduced is symmetric with respect to the energies of both quasi-degenerate states. For example, the non-symmetric transformations $E_i^{(0)},E_k^{(0)} \rightarrow E_1^{(0)}$ or $E_i^{(0)},E_k^{(0)} \rightarrow E_2^{(0)}$ would change the values of the off-diagonal matrix elements significantly.

Keeping this in mind, after rather tedious derivation one can obtain the calculation formulas for the second-order QED corrections due to the correlation effects in case of the quasi-degenerate states. The contribution of the diagram shown in Fig.~\ref{fig:2ph_2el}(a), which is referred to as the ladder (``ld'') diagram, is divided naturally into the reducible (``red'') and irreducible (``irr'') parts. The reducible term is determined by the conditions that the intermediate states $| n_1 n_2 \rangle$ belong to the $\Omega$ subspace, whereas the irreducible term corresponds to the remainder. The final expressions for the irreducible and reducible parts of the ladder contribution read as follows
\begin{align}
\label{eq:H2ik_ld_irr}
H^{{\rm ld,irr}}_{ik} = \frac{ F_i F_k }{ 2 } \, \sum_{P} (-1)^P 
   \sum_{n_1n_2}^{E_n^{(0)} \neq E_1^{(0)},E_2^{(0)}} 		
   \frac{i}{2\pi} \int_{-\infty}^\infty \! d\omega \,
&
 \Bigg\{\,
   \frac{ I_{Pi_1Pi_2n_1n_2}(\omega-\veps_{Pi_1}) I_{n_1n_2k_1k_2}(\veps_{k_1}-\omega) }
        { \big[ \omega - \veps_{n_1}(1-i0) \big] \big[ \bar{E}_{ik}^{(0)} - \omega - \veps_{n_2}(1-i0) \big] } \non 
&  \!      
 + \frac{ I_{Pi_1Pi_2n_1n_2}(\omega-\veps_{Pi_2}) I_{n_1n_2k_1k_2}(\veps_{k_2}-\omega) }
        {  \big[ \bar{E}_{ik}^{(0)} - \omega - \veps_{n_1}(1-i0) \big] \big[ \omega - \veps_{n_2}(1-i0) \big] } 
 \,\Bigg\} \, ,
\end{align}
\begin{align}
\label{eq:H2ik_ld_red}
H^{{\rm ld,red}}_{ik} = -\frac{ F_i F_k }{ 2 } \, \sum_{P} (-1)^P 
   \sum_{n_1n_2}^{E_n^{(0)} = E_1^{(0)},E_2^{(0)}}
   \frac{i}{2\pi} \int_{-\infty}^\infty \! d\omega \,
&
 \Bigg\{\,
   \frac{ I_{Pi_1Pi_2n_1n_2}(\omega-\veps_{Pi_1}) I_{n_1n_2k_1k_2}(\veps_{k_1}-\omega) }
        { \big[ \omega - \veps_{n_1} - i0 \big] \big[ \omega +  \veps_{n_2} - \bar{E}_{ik}^{(0)} - i0 \big] } \non 
&  \!   
 + \frac{ I_{Pi_1Pi_2n_1n_2}(\omega-\veps_{Pi_2}) I_{n_1n_2k_1k_2}(\veps_{k_2}-\omega) }
        { \big[ \omega +  \veps_{n_1} - \bar{E}_{ik}^{(0)} - i0 \big] \big[ \omega - \veps_{n_2} - i0 \big] }
 \,\Bigg\} \, .
\end{align}
The diagram in Fig.~\ref{fig:2ph_2el}(b) is termed as the crossed (``cr'') diagram. For its contribution we obtain the following formula
\begin{align}
\label{eq:H2ik_cr}
H^{{\rm cr}}_{ik} = \frac{ F_i F_k }{ 2 } \, \sum_{P} (-1)^P \sum_{n_1n_2}	
   \frac{i}{2\pi} \int_{-\infty}^\infty \! d\omega \,
&
 \Bigg\{\,
   \frac{ I_{Pi_1n_2n_1k_2}(\omega-\veps_{Pi_1}) I_{n_1Pi_2k_1n_2}(\veps_{k_1}-\omega) }
        { \big[ \omega - \veps_{n_1}(1-i0) \big] 
          \big[ \bar{E}_{ik}^{(0)} - \veps_{Pi_1} - \veps_{k_1} + \omega - \veps_{n_2}(1-i0) \big] } \non 
&  \!      
 +  \frac{ I_{Pi_1n_2n_1k_2}(\veps_{k_2}-\omega) I_{n_1Pi_2k_1n_2}(\omega-\veps_{Pi_2}) }
        { \big[ \bar{E}_{ik}^{(0)} - \veps_{Pi_2} - \veps_{k_2} + \omega - \veps_{n_1}(1-i0) \big] 
          \big[ \omega - \veps_{n_2}(1-i0) \big] } 
 \,\Bigg\} \, .
\end{align}
The formulas (\ref{eq:H2ik_ld_irr}) and (\ref{eq:H2ik_ld_red}) for the irreducible and reducible parts of the ladder term and the expression (\ref{eq:H2ik_cr}) for the crossed term differ from the ones presented in Ref.~\cite{Artemyev:2005:062104}. The final expressions corresponding to the diagrams in Fig.~\ref{fig:2ph_2el} have to be symmetric relative to the permutation of the electron lines, i.e., the transformation $i_1 \leftrightarrow i_2$, $k_1 \leftrightarrow k_2$, $n_1 \leftrightarrow n_2$ has to leave the total contribution to the matrix~$H$ unchanged. In Ref.~\cite{Artemyev:2005:062104} the symmetrization of the given expressions was implied, while in the present work the explicitly symmetric formulas are shown. 

Finally, in order to complete the discussion of the second-order QED corrections due to the interelectronic-interaction effects, it is worth noting that an additional minor modification for the off-diagonal elements of the matrix $H$ has been introduced in the present study compared to Ref.~\cite{Artemyev:2005:062104}. This modification influences the final result actually only in case of quasi-degenerate levels with $E_1^{(0)} \neq E_2^{(0)}$. As is known, the formula for the second order of the many-body perturbation theory (MBPT) can be obtained from the general second-order QED expression if one neglects the energy dependence of the photon propagator in the Coulomb gauge
and restricts the summation over the intermediate electron states to the positive-energy part of the Dirac spectrum.
Then, the $\omega$-integration can be carried out analytically employing Cauchy's residue theorem. Within this approximation, the contributions of the reducible~(\ref{eq:H2ik_ld_red}) and crossed~(\ref{eq:H2ik_cr}) terms vanish, while the irreducible part of the ladder diagram (\ref{eq:H2ik_ld_irr}) leads to the expression
\be 
\label{eq:H2ik_MBPT}
\tilde{H}^{(2)}_{ik}[{\rm MBPT}] = F_i F_k \, \sum_{P} (-1)^P 
   \psum_{n_1n_2}^{(+)} \,
   \frac{ I_{Pi_1Pi_2n_1n_2}(0) I_{n_1n_2k_1k_2}(0) }{ \bar{E}_{ik}^{(0)} - E_n^{(0)} } \, ,
\ee
where $E_n^{(0)} = \veps_{n_1} + \veps_{n_2}$ and the prime on the sum indicates that the terms with $E_n^{(0)} = E_1^{(0)},E_2^{(0)}$ have to be omitted in the summation. The formula~(\ref{eq:H2ik_MBPT}) for $E_i^{(0)} \neq E_k^{(0)}$ differs slightly from the standard MBPT expression, which can be derived from the irreducible contribution without the replacement $E_i^{(0)},E_k^{(0)} \rightarrow \bar{E}_{ik}^{(0)}$ introduced,
\begin{align}
\label{eq:H2ik_MBPT_rigorous}
H^{(2)}_{ik}&[{\rm MBPT}] =
   \frac{ F_i F_k }{2} \, \sum_{P} (-1)^P 
   \psum_{n_1n_2}^{(+)} \,
   \left[ 
   \frac{1}{E_i^{(0)} -  E_n^{(0)}} + \frac{1}{E_k^{(0)} -  E_n^{(0)}} 
   \right] 
   I_{Pi_1Pi_2n_1n_2}(0) I_{n_1n_2k_1k_2}(0)   \nonumber \\
&= 
   F_i F_k \, \sum_{P} (-1)^P 
   \psum_{n_1n_2}^{(+)} \,
   \frac{ \bar{E}_{ik}^{(0)} -  E_n^{(0)} }
        { \big( E_i^{(0)} -  E_n^{(0)} \big)\big( E_k^{(0)} -  E_n^{(0)} \big) } \, 
   I_{Pi_1Pi_2n_1n_2}(0) I_{n_1n_2k_1k_2}(0) \, .
\end{align}
Once again, we stress that the Coulomb gauge is implied for the photon propagator in Eqs.~(\ref{eq:H2ik_MBPT}) and (\ref{eq:H2ik_MBPT_rigorous}).
Defining in a self-consistent way the ``pure'' QED correction to the matrix $H$ by
\be
\label{eq:H2_nondiag_pureQED}
H^{(2)}_{ik}[{\rm QED}] \equiv \tilde{H}_{ik}^{(2)} - \tilde{H}^{(2)}_{ik}[{\rm MBPT}] \, ,
\ee
where
\be 
\label{eq:H2_ik}
\tilde{H}_{ik}^{(2)} = H^{{\rm ld,irr}}_{ik} + H^{{\rm ld,red}}_{ik} + H^{{\rm cr}}_{ik} \, 
\ee
is the sum of Eqs.~(\ref{eq:H2ik_ld_irr}), (\ref{eq:H2ik_ld_red}), and (\ref{eq:H2ik_cr}),
it is natural to consider the expression
\be
\label{eq:H2_ik_total}
H^{(2)}_{ik} \equiv H^{(2)}_{ik}[{\rm QED}] + H^{(2)}_{ik}[{\rm MBPT}]
  = \tilde{H}_{ik}^{(2)} - \tilde{H}^{(2)}_{ik}[{\rm MBPT}] + H^{(2)}_{ik}[{\rm MBPT}] \, ,
\ee
as the final second-order contribution to the matrix $H$. In contrast to Eq.~(\ref{eq:H2_ik}), the expression (\ref{eq:H2_ik_total}) leads to the standard formula for the second-order contribution within the Breit approximation. This modification distinguishes the present consideration of the second-order interelectronic-interaction effects from the calculations performed in Ref.~\cite{Artemyev:2005:062104}. 

\subsection{Higher-order correlation effects \label{sec:1:2}}

High-precision calculations of energy levels in few-electron ions have to take into account also the higher-order correlation effects at least within the lowest-order relativistic approximation. In Ref.~\cite{Artemyev:2005:062104}, the interelectronic-interaction contributions due to the exchange by three or more photons were included by employing the results of the $1/Z$ expansions performed within the $LS$-coupling scheme. The corresponding coefficients for nonrelativistic energies were taken from Refs.~\cite{Sanders:1969:84,Aashamar:1970:3324}, while for the relativistic Breit-Pauli correction the results from Ref.~\cite{Drake:1988:586} were used. In case of the quasi-degenerate levels $(1s\,2p_{1/2})_1$ and $(1s\,2p_{3/2})_1$, the $jj$-$LS$ recoupling matrix $R$, defined according to
\be
\label{eq:R_recoupl}
\dvec{ | \, 1s2p{\,}^3P_1  \, \rangle }{ | \, 1s2p{\,}^1P_1 \, \rangle } = R \,
\dvec{ | \, (1s\,2p_{1/2})_1 \, \rangle }{ | \, (1s\,2p_{3/2})_1 \, \rangle }, \qquad
R = \frac{1}{\sqrt{3}} \begin{pmatrix} \sqrt{2} & -1 \\ 1 & \sqrt{2} \end{pmatrix} \, ,
\nonumber
\ee
was employed to obtain the contribution of the higher-order correlation effects to the matrix $H$ in the $jj$ coupling (see the discussion in Refs.~\cite{Artemyev:2005:062104, Drake:1988:586}). The matrix $R$ relates the wave functions constructed in two different couplings within the nonrelativistic approximation. Nevertheless, it is highly desirable to have a kind of self-consistent procedure for consideration of the correlation effects which treats all orders of the perturbation theory within the relativistic approach on equal footing.
Additional motivation for developing an alternative procedure is that the method employed in Ref.~\cite{Artemyev:2005:062104} does not allow for the calculations within the extended Furry picture. 
The latter implies modification of the zeroth-order approximation by including a local screening potential into the Dirac Hamiltonian $h^{\rm D}$ in Eq.~(\ref{eq:dirac}). On the one hand, this method was found to be very useful in the QED calculations of the different atomic properties in few-electron ions \cite{Sapirstein:2001:022502, Sapirstein:2001:032506, Glazov:2006:330, Artemyev:2007:173004, Yerokhin:2007:062501, Kozhedub:2010:042513, Sapirstein:2011:012504, Volotka:2014:253004, Sapirstein:2015:062508, Aleksandrov:2018:062521, Malyshev:2019:010501_R} and many-electron atoms~\cite{Sapirstein:2002:042501, Chen:2006:042510, Sapirstein:2003:022512, Sapirstein:2006:042513}, but, on the other hand it leads to the rearrangement of all perturbation series. 
Since the $1/Z$-expansion coefficients are known for the Coulomb potential of the point nucleus only, the related calculations with another choice of the initial approximation are not possible. 

In the present work we employ two independent methods in order to evaluate the higher-order correlation effects. Both methods use the Dirac-Coulomb-Breit (DCB) Hamiltonian to treat the interelectronic interaction,
\be
\label{eq:DCB}
H_{\rm DCB} &=& \Lambda^{(+)} \left[ H_0 + V_{\rm int} \right] \Lambda^{(+)} \, ,
		\label{eq:H_DCB} \\[1mm]
H_0 &=& \sum_i^N h^{\rm D}_i \, , \qquad 
V_{\rm int} = \sum_{i<j}^N \left[ V^{\rm C}_{ij} + V^{\rm B}_{ij} \right]
\\[1mm]		
h^{\rm D} &=& \balpha \cdot \bm{p} + \beta m + V_{\rm nucl} \, , \label{eq:h_Dirac} \\[1mm]
V^{\rm C}_{ij} &=& \frac{\alpha}{r_{ij}} \, , \qquad
V^{\rm B}_{ij} = -\frac{\alpha}{2r_{ij}} \left[ \balpha_i \cdot \balpha_j + \frac{(\balpha_i \cdot \br_{ij})(\balpha_j \cdot \br_{ij})}{r_{ij}^2} \right] \, ,
\ee
where $\br_{ij}=\br_i-\br_j$, $r_{ij}=|\br_{ij}|$, and $V^{\rm C}$ and  $V^{\rm B}$ are the Coulomb and Breit parts of the electron-electron interaction operator within the Breit approximation. One can note that $V^{\rm C}_{ij} + V^{\rm B}_{ij} = e^2 \alpha^\mu_i \alpha^\nu_j D_{\mu\nu}(0,\br_{ij})$ provided the photon propagator $D_{\mu\nu}$ in the Coulomb gauge is considered. In Eq.~(\ref{eq:DCB}), $\Lambda^{(+)}$ is the product of the one-electron positive-energy-states projectors corresponding to the potential $V_{\rm nucl}$.
The generalization of the Hamiltonian~(\ref{eq:DCB}) to case of the extended Furry picture is discussed in details, e.g., in Ref.~\cite{Kaygorodov:2019:032505}. 
The key point for merging the \textit{ab initio} QED results with the higher-order interelectronic-interaction contributions is that the projectors~$\Lambda^{(+)}$ must be defined with respect to the same Dirac Hamiltonian $h^{\rm D}$ in Eq.~(\ref{eq:dirac}) which provides the initial approximation for the QED perturbation theory.
Therefore, $h^{\rm D}$ and $\Lambda^{(+)}$ should be defined consistently when the QED calculations within the extended Furry picture are performed. As indicated, e.g., in Refs.~\cite{Kaygorodov:2019:032505}, 
this is not generally needed in relativistic calculations based on the DCB Hamiltonian. 


The first approach employed in the present work for calculations of the higher-order correlation effects is the large-scale configuration interaction (CI) method in the basis of the Dirac-Sturm (DS) orbitals~\cite{Bratzev:1977:173, Tupitsyn:2003:022511}.
In case of single levels, the procedure to extract the desired third- and higher-order contributions from the total CI results is well known. In order to subtract the zeroth-, first-, and second-order terms, one can calculate them within the Breit approximation by the perturbation theory using the same basis set. Alternatively, the corresponding terms can be obtained by evaluating the derivatives of the CI energies with respect to the factor artificially introduced before the interaction term $V_{\rm int}$ in the DCB Hamiltonian (see the details, e.g., in Refs.~\cite{Kozhedub:2010:042513, Malyshev:2017:022512}).

In order to obtain the contribution of the higher-order correlation effects to the matrix $H$ for the set of $s$ (quasi-)degenerate levels within the CI approach, we have developed the following procedure which was applied first in Ref.~\cite{Malyshev:2019:010501_R}.
The CI method allows one to calculate the energies $E_1^{\rm CI},\ldots,E_s^{\rm CI}$ and the corresponding many-electron wave functions $\left\{\Psi_j\right\}_{j=1}^s$ for the (quasi-)degenerate states. Therefore, in the spirit of the TTGF method, one can introduce the projection of the CI Green's function on the subspace $\Omega$ spanned by the unperturbed wave functions $\left\{u_j\right\}_{j=1}^s$,
\be
\label{eq:Green_CI}
g^{\rm CI}(E) = P^{(0)} \left[ \sum_{j=1}^s \frac{| \Psi_j \rangle \langle \Psi_j | }{E-E_j^{\rm CI}} + \Xi \right] P^{(0)} \, ,
\ee
where the projector $P^{(0)}$ is defined in Eq.~(\ref{eq:quasi_P0}) and the term $\Xi$ includes the remaining part of the many-electron CI spectrum. The CI versions of the operators (\ref{eq:TTGF:K}) and (\ref{eq:TTGF:P}) can be determined by
\be
\label{eq:TTGF:K_CI}
\hat{K}^{\rm CI} &\equiv& \frac{1}{2\pi i} \oint_\Gamma \! dE \,\, E g^{\rm CI}(E) \, , \\[2mm]
\label{eq:TTGF:P_CI}
\hat{P}^{\rm CI} &\equiv& \frac{1}{2\pi i} \oint_\Gamma \! dE \,\,   g^{\rm CI}(E) \, ,
\ee
where the contour $\Gamma$ surrounds all the poles corresponding to the CI energies $E_1^{\rm CI},\ldots,E_s^{\rm CI}$ and keeps outside all the other singularities arising from the term $\Xi$. Substituting $g^{\rm CI}(E)$ from Eq.~(\ref{eq:Green_CI}) into Eqs.~(\ref{eq:TTGF:K_CI}) and (\ref{eq:TTGF:P_CI}) we obtain the following $s\times s$ matrices
\be
\label{eq:K_ik_CI}
K_{ik}^{\rm CI} &  =  & \sum_{j=1}^s E_j^{\rm CI} \, \langle u_i | \Psi_j \rangle \langle \Psi_j | u_k \rangle \, , \\
\label{eq:P_ik_CI}
P_{ik}^{\rm CI} &  =  & \sum_{j=1}^s              \, \langle u_i | \Psi_j \rangle \langle \Psi_j | u_k \rangle \, .
\ee
Finally, the matrix $H$ can be constructed from the matrices~(\ref{eq:K_ik_CI}) and (\ref{eq:P_ik_CI}) according to Eq.~(\ref{eq:TTGF:H}). The described method allows one to take into account the interelectronic-interaction effects to all orders in $1/Z$. The low-order terms can be subtracted using the PT calculations performed within the Breit approximation with the same basis set. For two-electron ions, the zeroth- and second-order contributions to the matrix $H$ within the Breit approximation are presented in Eqs.~(\ref{eq:H0_ik}) and (\ref{eq:H2ik_MBPT_rigorous}), respectively. The first-order term is given by Eq.~(\ref{eq:H1_ik}) considered in the Coulomb gauge with $\Delta_{1,2}$ replaced with zero. 

The second method employed is based on the perturbation theory in the basis of Slater determinants. The latter are constructed from the one-electron wave functions of the finite basis set, which is exactly the same as the one used for the QED calculations. In order to access arbitrary high orders of PT we employ the recursive formulation. This approach first implemented in Ref.~\cite{Glazov:2017:46} was used in the calculations of the higher-order interelectronic-interaction contributions to the energies of boronlike ions \cite{Malyshev:2017:103, Malyshev:2017:022512}. Extended to the case of multiple perturbations, it has been applied recently to the Zeeman splitting in lithiumlike and boronlike ions \cite{Varentsova:2018:043402,Arapoglou:2019:253001,Glazov:2019}, including the nuclear recoil effect \cite{Shabaev:2017:263001,Shabaev:2018:032512,Aleksandrov:2018:062521}. In the form presented in  Ref.~\cite{Glazov:2017:46}, it has a limitation that the reference state in the zeroth-order approximation must be represented exactly by one Slater determinant. For the case of excited states in heliumlike ions it has been trivially generalized to deal with the many-determinant reference states. This generalization is closely connected to the non-trivial construction of the PT for the (quasi-)degenerate states. Since pioneering works by Kato \cite{Kato:1949:note} and Bloch \cite{Bloch:1958:329} considerable effort has been devoted to development of the general PT expressions for the cases of degenerate and quasi-degenerate states (see, e.g., \cite{Kato:1966:book:eng,Killingbeck:1977:963,Reed:1978:book:eng} and references therein). The compact recursive algorithm for derivation of these expressions has been proposed recently by Brouder and coauthors~\cite{Brouder:2012:2256}. 
Within the recursive scheme, the $p$-th order contributions to the energies and wave functions of the reference state(s) are constructed from the lower-order contributions, from zeroth to $(p-1)$-th~\cite{Glazov:2017:46}. For the (quasi-)degenerate states the corresponding expression can be written for the matrix element $H_{ik}^{(p)}$. The most problematic question is which one of the zeroth-order energies $\{E_j^{(0)}\}_{j=1}^s$ appears in the denominator in each particular term. The answer to this question has been implemented within the combinatorial algorithm given in Ref.~\cite{Brouder:2012:2256}. While the complete set of formulas cannot be written in a compact form, we illustrate the PT for the (quasi-)degenerate states by the closed formula for the third-order contribution,
\be
\label{eq:H3_ik}
  H_{ik}^{(3)}[{\rm MBPT}] =
  \frac{1}{2} \left\{
   \psum_{nm}^{(+)}
   \frac{\matrixel{u_i}{V_{\rm int}}{\Psi_n^{(0)}}
         \matrixel{\Psi_n^{(0)}}{V_{\rm int}}{\Psi_m^{(0)}}
         \matrixel{\Psi_m^{(0)}}{V_{\rm int}}{u_k}}
   {\big(E_k^{(0)} - E_n^{(0)}\big)\big(E_k^{(0)} - E_m^{(0)}\big)}
 \right.
\nonumber\\
 \left.
 - \psum_{n}^{(+)} \, \sum_{j=1}^{s}
   \frac{\matrixel{u_i}{V_{\rm int}}{\Psi_n^{(0)}}
         \matrixel{\Psi_n^{(0)}}{V_{\rm int}}{u_j}
         \matrixel{u_j}{V_{\rm int}}{u_k\vphantom{\Psi_n^{(0)}}}}
   {\big(E_k^{(0)} - E_n^{(0)}\big)\big(E_j^{(0)} - E_n^{(0)}\big)}
 + (i \leftrightarrow k)
 \right\}
\,.
\ee
Here, $\ket{\Psi_n^{(0)}}$ and $\ket{\Psi_m^{(0)}}$ are the unperturbed wave functions orthogonal to the $\Omega$ subspace, which is stressed by the prime over the sum symbol. The symmetrization is made explicitly by averaging with the transposed expression $(i \leftrightarrow k)$. For the two-electron systems under consideration this formula can be rewritten in the following form
\be
\label{eq:H3_ik_2el}
  H_{ik}^{(3)}[{\rm MBPT}] =
   \frac{ F_i F_k }{2} \, \sum_{P} (-1)^P \, \psum_{n_1n_2}^{(+)} 
   \left\{ \,
   \psum_{m_1m_2}^{(+)}
   \frac{I_{Pi_1Pi_2n_1n_2}(0) I_{n_1n_2m_1m_2}(0) I_{m_1m_2k_1k_2}(0)}
   {(\veps_{k_1} + \veps_{k_2} - \veps_{n_1} - \veps_{n_2})(\veps_{k_1} + \veps_{k_2} - \veps_{m_1} - \veps_{m_2})}
 \right.
\nonumber\\
 \left.
 - \sum_{Q} (-1)^Q \, \sum_{j=1,2}
   \frac{I_{Pi_1Pi_2n_1n_2}(0) I_{n_1n_2Qj_1Qj_2}(0) I_{Qj_1Qj_2k_1k_2}(0)}
   {(\veps_{k_1} + \veps_{k_2} - \veps_{n_1} - \veps_{n_2})(\veps_{j_1} + \veps_{j_2} - \veps_{n_1} - \veps_{n_2})}
 + (i \leftrightarrow k)
 \right\}
\,.
\ee
Second permutation operator $Q$ is introduced here, and all other notations have been defined before.
We note that, in contrast to the CI method, the recursive PT provides direct access to the required contributions, specifically, to the third and higher orders in the present case. No subtraction of the leading-order terms is needed, which is advantageous from the numerical point of view.
%


\section{Numerical results and discussion \label{sec:2}}

\subsection{Electron-electron correlation effects in He-like ions \label{sec:2:1}}

In this section, we present the results of the calculations of the correlation effects in heliumlike ions. For the single  $(1s\,1s)_0$, $(1s\,2s)_0$, $(1s\,2s)_1$, $(1s\,2p_{1/2})_0$, and $(1s\,2p_{3/2})_2$ states, the interelectronic-interaction contributions to the binding energies are evaluated. For the pair of quasi-degenerate levels $(1s\,2p_{1/2})_1$ and $(1s\,2p_{3/2})_1$, the corresponding contributions to the $2\times 2$ matrix $H$ are obtained. In what follows, when we refer to the mixing configurations, $(1s\,2p_{1/2})_1$ and $(1s\,2p_{3/2})_1$ stand for the diagonal contributions while ``off-diag.'' denotes the contributions to the off-diagonal matrix elements. In the calculations, the Fermi model with the thickness parameter equal to $2.3$~fm is employed in order to describe the nuclear charge distribution. The root-mean-square (rms) radii are taken from Ref.~\cite{Angeli:2013:69}. The CODATA 2014 recommended values of the fundamental constants~\cite{Mohr:2016:035009} are used: $\alpha^{-1}=137.035\,999\,139(31)$ and $mc^2=0.510\,998\,9461(31)$~MeV.

We start with the results obtained within the rigorous QED approach. The electron-correlation contributions corresponding to the two-photon exchange diagrams depicted in Fig.~\ref{fig:2ph_2el} are presented for the single and quasi-degenerate states in Tables~\ref{tab:2ph_QED_single} and \ref{tab:2ph_QED_quasi}, respectively. The QED calculations of the second-order interelectronic-interaction effects for the $n=1$ and $n=2$ levels in heliumlike ions are based on Eqs.~(\ref{eq:H2ik_ld_irr})--(\ref{eq:H2ik_cr}). The summation over the intermediate electronic states is performed by using the finite-basis set which is constructed from B-splines~\cite{Sapirstein:1996:5213} within the framework of the dual-kinetic-balance method~\cite{splines:DKB}. The integration over $\omega$ is carried out numerically after applying Wick's rotation. The uncertainties indicated in the parentheses are due to the numerical computation errors. The off-diagonal matrix elements evaluated according to Eqs.~(\ref{eq:H2_ik}) and (\ref{eq:H2_ik_total}) are shown in the fourth and fifth columns of Table~\ref{tab:2ph_QED_quasi}, respectively. One can see that the discrepancy between the results tends to zero with decreasing the nuclear charge number $Z$. On the other side, the difference becomes noticeable for high-$Z$ heliumlike ions. As mentioned in Sec.~\ref{sec:1:1}, the deviation between Eqs.~(\ref{eq:H2_ik}) and (\ref{eq:H2_ik_total}) is at the level of the higher-order QED effects which are beyond the scope of the present study. Indeed, this difference turns to be well within our estimation of the uncertainty due to the uncalculated QED contributions of the third and higher orders in $1/Z$ (see the discussion below). 

For the single states, the two-photon exchange diagrams represent the gauge invariant set. However, this is not the case for the mixing $(1s2p)_1$ levels. The individual contributions to the matrix $H$ of a particular order in $1/Z$ may vary from gauge to gauge. Only the eigenvalues of the complete matrix $H$ evaluated to all orders in $\alpha$ and $1/Z$ are gauge invariant. In the present study, we are interested in merging the QED results with the higher-order contributions obtained within the Breit approximation. As noted above, the latter calculations are based on the DCB Hamiltonian~(\ref{eq:DCB}) which, in turn, is naturally related to the Coulomb gauge~\cite{Shabaev:1993:4703}. For this reason, the second-order interelectronic-interaction contributions to the matrix $H$ presented in Table~\ref{tab:2ph_QED_quasi} were obtained by operating in the Coulomb gauge. However, as a check of the numerical procedure, we repeated the calculations of the two-photon exchange for all the states in the Feynman gauge as well. For the single levels, the results in both gauges were found in excellent agreement with each other. For the mixing configurations, the difference between the second-order contributions calculated in the Coulomb and Feynman gauges lies within the uncertainties specified in Table~\ref{tab:2ph_QED_quasi}. For low-$Z$ ions, the difference is negligible. It increases with $Z$ and, for heliumlike uranium, its absolute value reaches approximately 0.14~meV and 0.02~meV for the diagonal and off-diagonal matrix elements, respectively. 

In Tables~\ref{tab:2ph_QED_single} and \ref{tab:2ph_QED_quasi}, our results for the two-photon exchange correction for the $n=1$ and $n=2$ states in heliumlike ions are compared with the previous QED calculations~\cite{Blundell:1993:2615, Mohr:2000:052501, Andreev:2001:042513, Asen:2002:032516, Andreev:2003:012503, Andreev:2004:062505, Artemyev:2005:062104}. It is seen that the obtained results are generally in agreement with the values available in the literature but have higher accuracy. As indicated in Ref.~\cite{Malyshev:2019:010501_R}, a small discrepancy with the results presented in Ref.~\cite{Artemyev:2005:062104} takes place for the $J=0$ states, namely, for $(1s\,1s)_0$, $(1s\,2s)_0$, and $(1s\,2p_{1/2})_0$. It is most pronounced for the ground state. The reason for the discrepancy is probably in the underestimation of the uncertainty of the calculations performed in Ref.~\cite{Artemyev:2005:062104}.

\begin{figure}
\centering

{}
\includegraphics[width=0.48\textwidth]{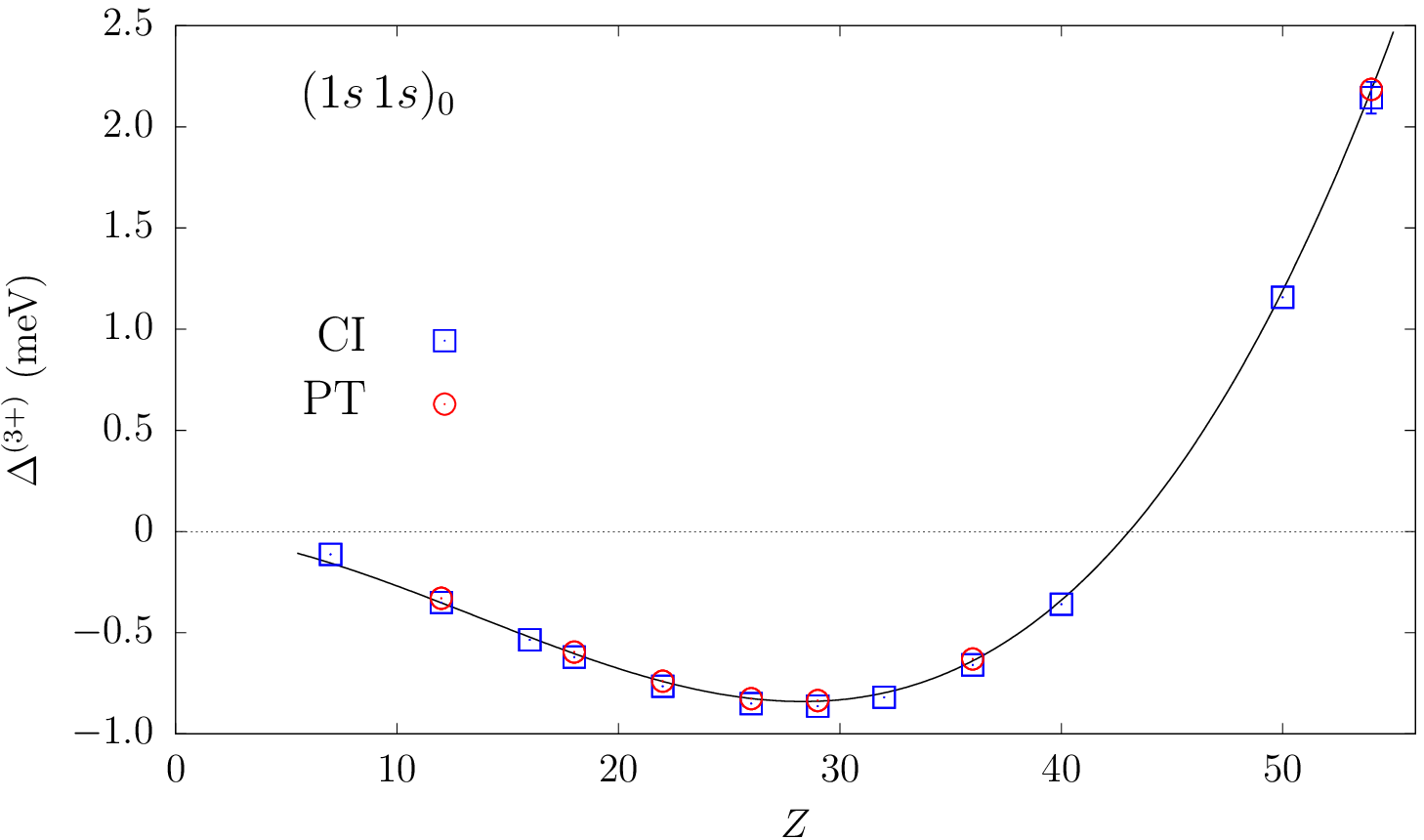}
{}
\hfill
{}
\includegraphics[width=0.48\textwidth]{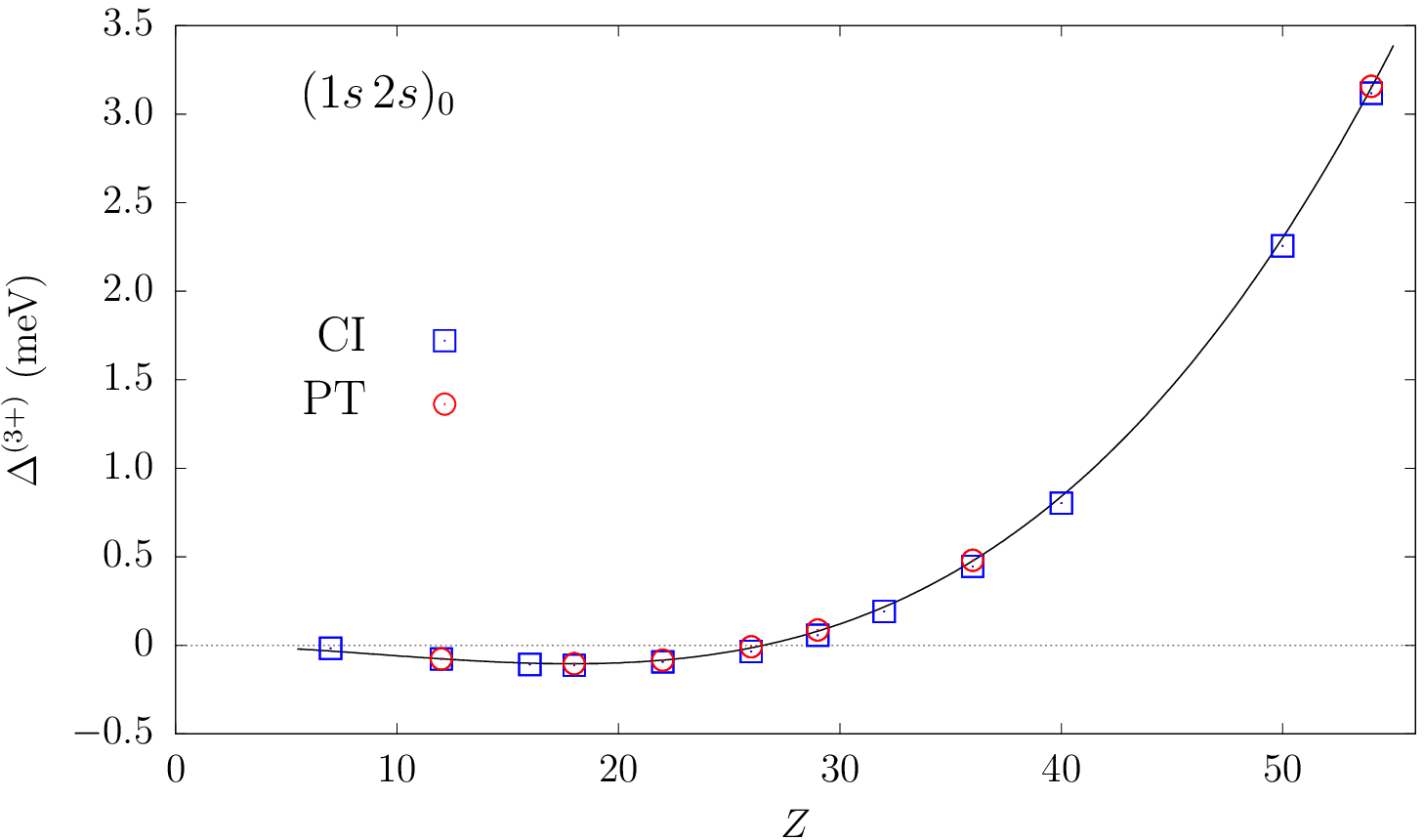}
{}

\bigskip

{}
\includegraphics[width=0.48\textwidth]{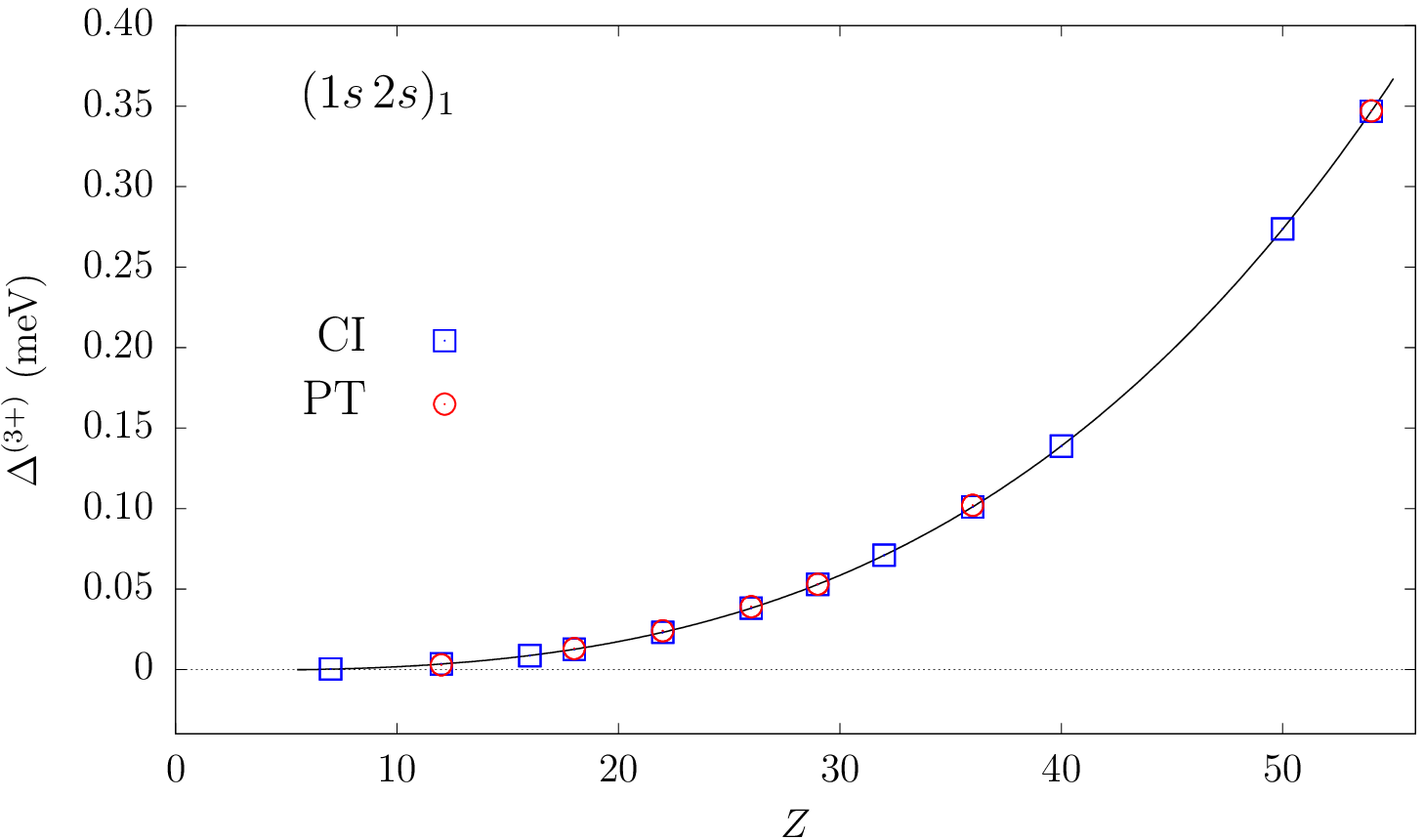}
{}
\hfill
{}
\includegraphics[width=0.48\textwidth]{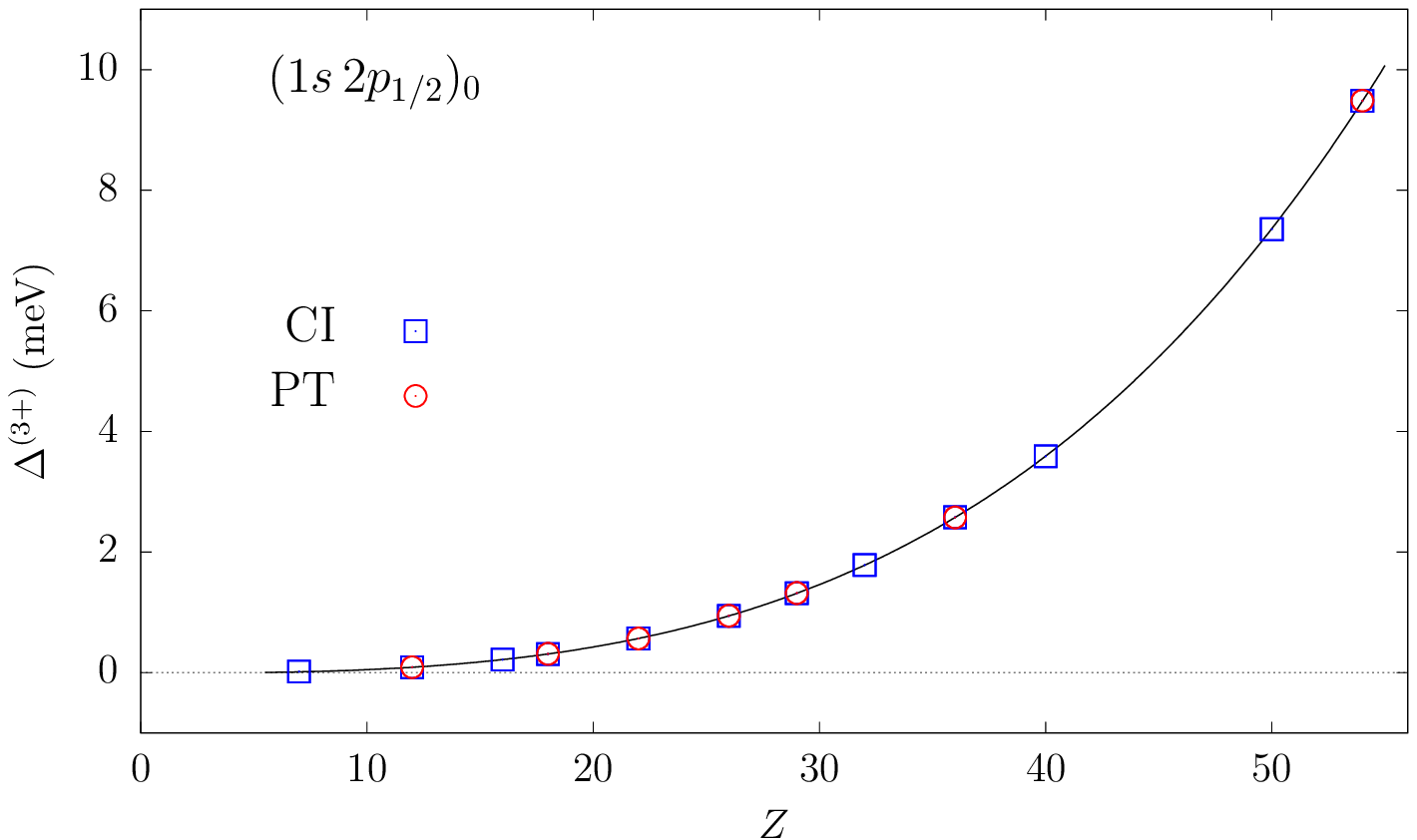}
{}

\bigskip

\includegraphics[width=0.48\textwidth]{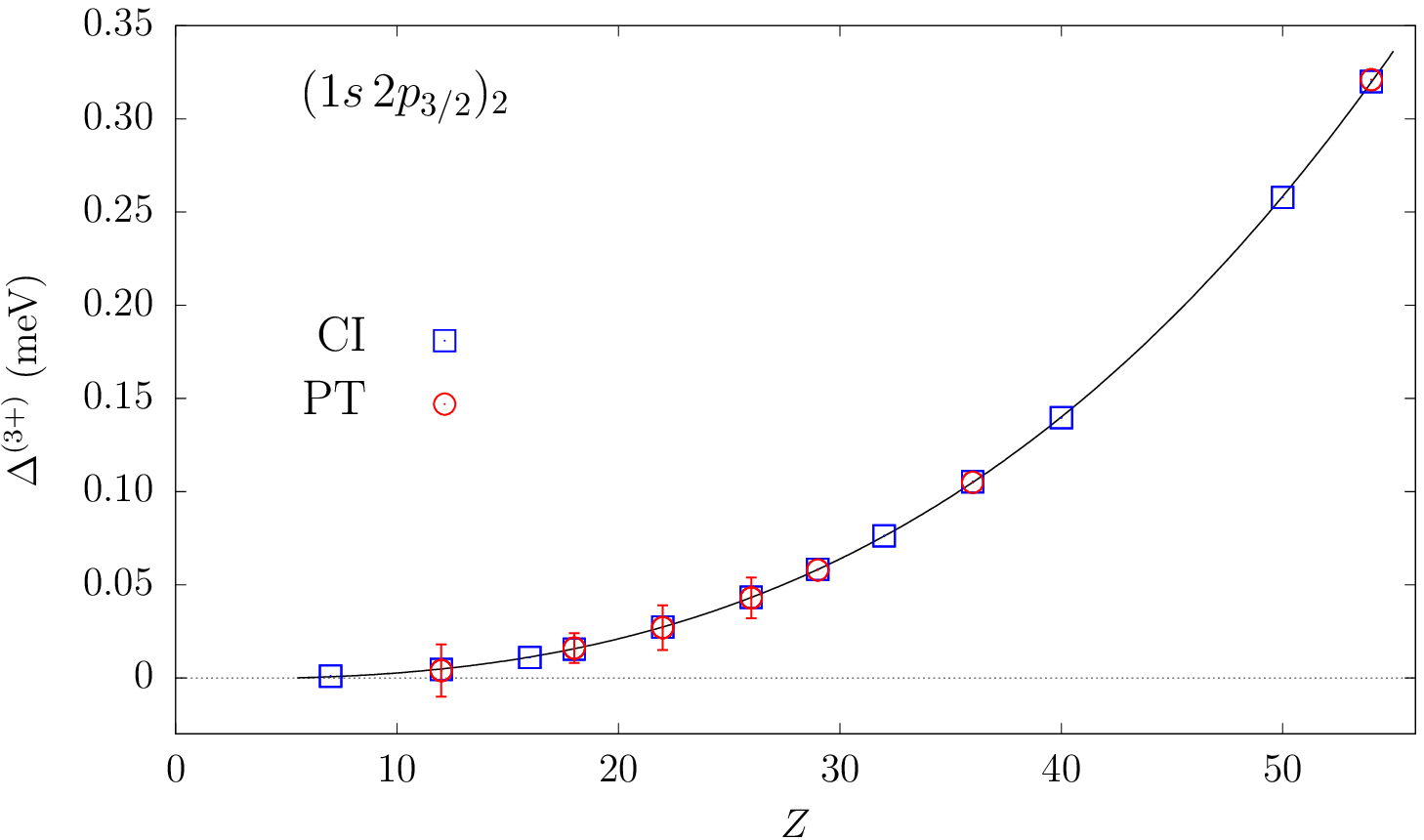}

\caption{\label{fig:3plus} Comparison of the different calculations within the Breit approximation of the third- and higher-order interelectronic-interaction contributions to the binding energies of the $n=1$ and $n=2$ single states in He-like ions, in meV. The difference $\Delta^{(3+)}$ between our results and the values obtained within the $1/Z$ expansion is plotted. The results of the CI approach are shown as blue squares, while the results of the PT calculations are depicted as red circles. See the text for details.}
\end{figure}

\begin{figure}
\centering

{}
\includegraphics[width=0.48\textwidth]{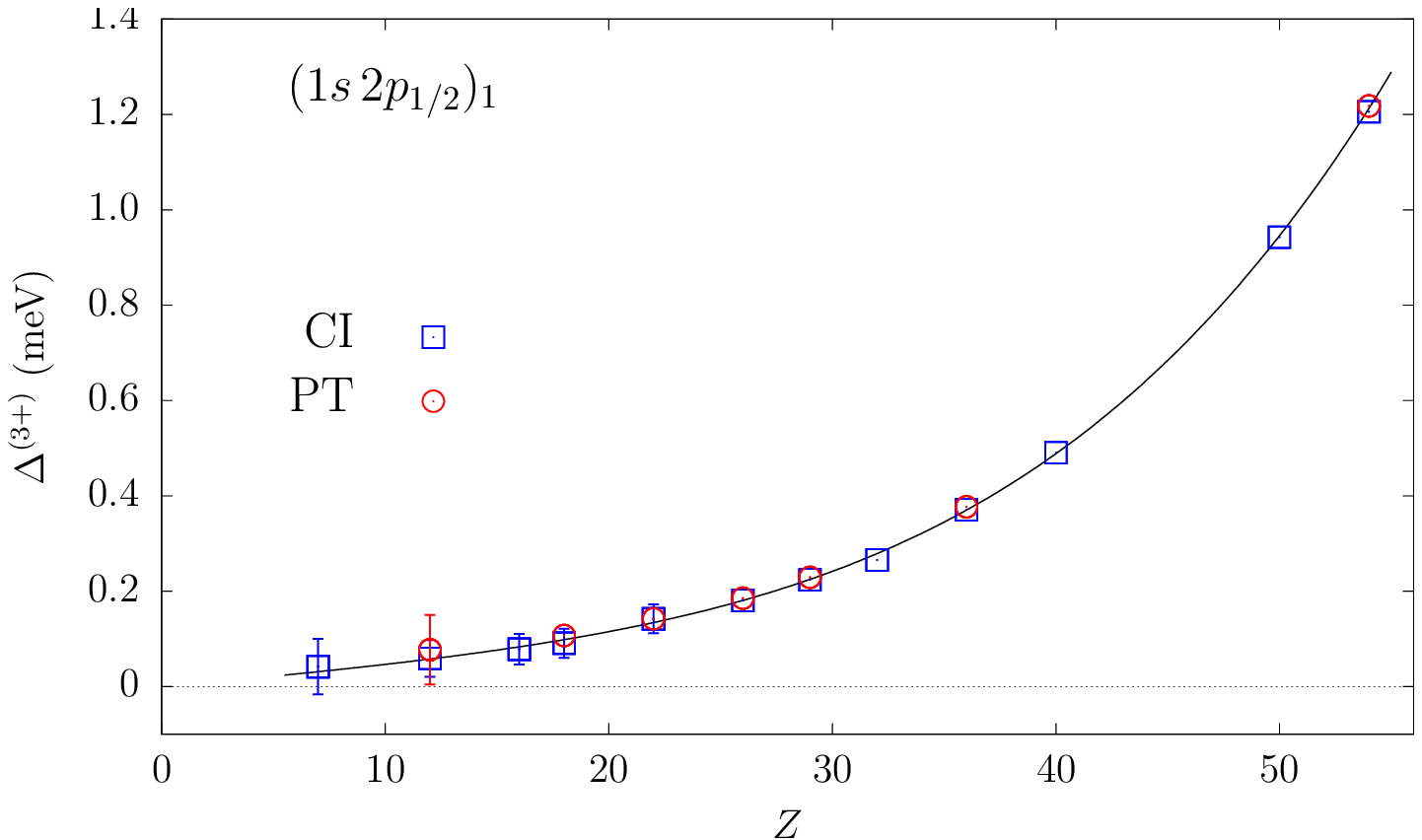}
{}
\hfill
{}
\includegraphics[width=0.48\textwidth]{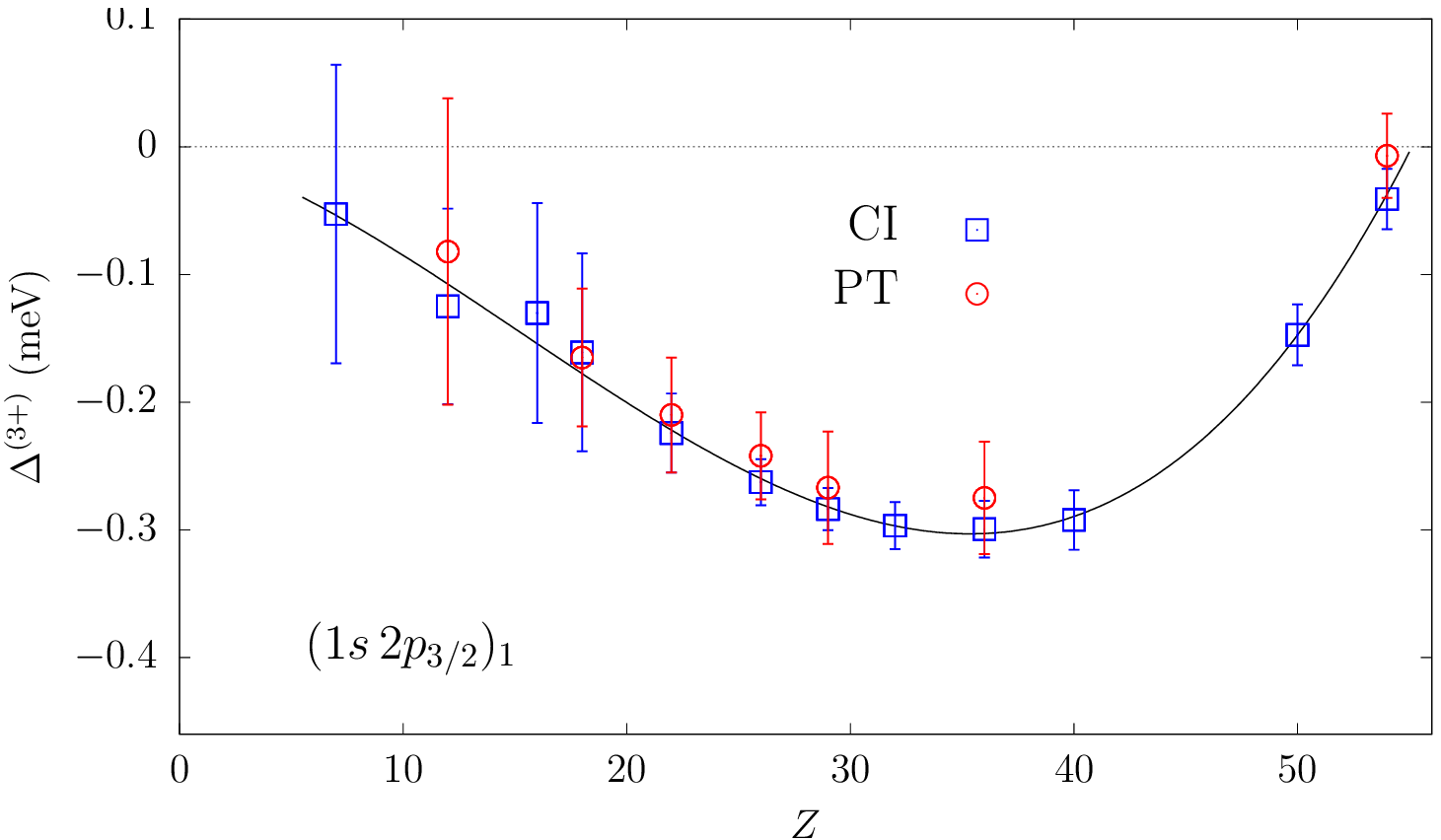}
{}

\bigskip

\includegraphics*[width=0.48\textwidth]{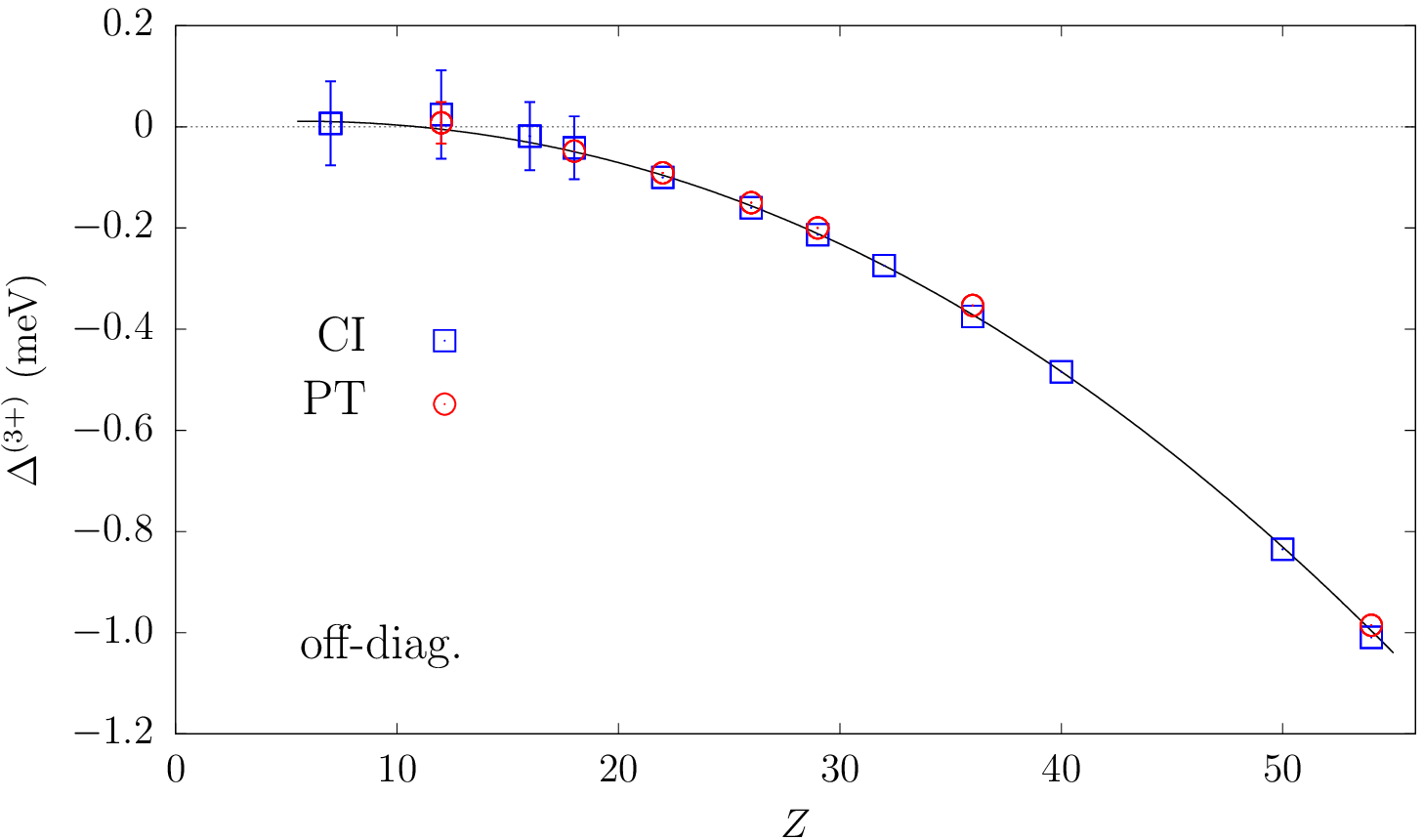}

\caption{\label{fig:3plus_quasi} Comparison of the different calculations within the Breit approximation of the third- and higher-order interelectronic-interaction contributions to the matrix $H_{ik}$ for the $n=2$ quasi-degenerate states in He-like ions, in meV. The difference $\Delta^{(3+)}$ between our results and the values obtained within the $1/Z$ expansion is plotted. $(1s\,2p_{1/2})_1$ and $(1s\,2p_{3/2})_1$ stand for the corresponding diagonal matrix elements of the operator~$H$, whereas ``off-diag.'' corresponds to the off-diagonal matrix elements. The results of the CI approach are shown as blue squares, while the results of the PT calculations are depicted as red circles. See the text for details.}
\end{figure}

The second key point of the present calculations is the evaluation of the third- and higher-order correlation effects within the Breit approximation. As a benchmark, we consider the approach which was employed in Refs.~\cite{Artemyev:2005:062104, Drake:1988:586}. This approach is based on the $1/Z$ expansions of the nonrelativistic energies and the Breit-Pauli correction. In order to evaluate the desired contribution in the framework of this method, we used the $1/Z$-expansion coefficients tabulated in Ref.~\cite{Yerokhin:2010:022507}. Since the off-diagonal matrix element of the Breit-Pauli Hamiltonian between the $| \, 1s2p{\,}^3P_1  \, \rangle$ and $| \, 1s2p{\,}^1P_1  \, \rangle$ states cannot be unambiguously identified from Ref.~\cite{Yerokhin:2010:022507}, we employed the corresponding coefficients from Ref.~\cite{Drake:1988:586}. Our results for the third- and higher-order interelectronic-interaction contributions for the single and quasi-degenerate levels in heliumlike ions are presented in Figs.~\ref{fig:3plus} and \ref{fig:3plus_quasi}, respectively. In view of the high accuracy of the calculations, it is convenient to analyze the difference $\Delta^{(3+)}$ between our results and the values obtained within the $1/Z$-expansion method. The results of the CI approach in terms of $\Delta^{(3+)}$ are shown in Figs.~\ref{fig:3plus} and \ref{fig:3plus_quasi} as blue squares, and the corresponding PT results are given as red circles. For each particular state and each nuclear charge, we perform our calculations with a wide variety of the basis sets (constructed from the DS orbitals and the B-splines in the CI and PT methods, respectively). By increasing the size of the basis in all possible directions and analyzing the successive increments of the results, we obtain a reliable estimation of how well our CI and PT calculations converge. In Figs.~\ref{fig:3plus} and \ref{fig:3plus_quasi}, the uncertainties of the calculated higher-order correlation effects are shown only if they exceed the size of the squares or circles plotted. One can see that the results of both independent approaches are in good agreement with each other. The deviation from the $1/Z$-expansion values, $\Delta^{(3+)}$, arises from the different treatment of the relativistic effects. While in Refs.~\cite{Drake:1988:586,Yerokhin:2010:022507} the Breit part $V^{\rm B}$ of the electron interaction operator is considered as a perturbation to first order, it is taken into account to all orders in $1/Z$ in the present calculations. At the $\alpha Z\rightarrow 0$ limit, the deviation tends to zero, as it should be.


Finally, all the interelectronic-interaction corrections for the single and quasi-degenerate states in heliumlike iron ($Z=26$), xenon ($Z=54$), and uranium ($Z=92$) are collected in Tables~\ref{tab:total_single} and \ref{tab:total_quasi}. For each ion, the first line contains the zeroth-order approximation $E_{\rm Dirac}^{(0)}$ arising from the Dirac equation~(\ref{eq:dirac}), that is the sum of the Dirac energies with the rest masses subtracted. The interelectronic-interaction contributions of the first, second, and higher orders evaluated within the Breit approximation are given in the rows $E_{\rm Breit}^{(1)}$, $E_{\rm Breit}^{(2)}$, and $E_{\rm Breit}^{(3+)}$, respectively. The first- and second-order QED corrections, $E_{\rm QED}^{(1)}$ and $E_{\rm QED}^{(2)}$, are obtained as the differences between the contributions of the one- and two-photon exchange diagrams calculated within the rigorous QED approach and within the Breit approximation. The uncertainty $E^{(3+)}_{\rm QED}$ due to uncalculated QED contributions to the higher-order interelectronic-interaction effects is conservatively estimated as $E^{(3+)}_{\rm Breit}$ multiplied by $2\,E^{(2)}_{\rm QED}/E^{(2)}_{\rm Breit}$. This uncertainty calculated for the ground state is used for all the other states also in order to avoid an underestimation due to anomalously small values of the $E^{(3+)}_{\rm Breit}$ contribution for some states and ions. As can be seen, the estimation employed covers the difference between the calculations of the second-order contribution for the off-diagonal matrix element by Eqs.~(\ref{eq:H2_ik}) and (\ref{eq:H2_ik_total}). The total interelectronic-interaction correction to the binding energies of the single states and to the matrix $H$ for the quasi-degenerate states is shown in line labeled as $E_{\rm int}$. The last line represents the sum of the zeroth-order contribution and the total interelectronic-interaction correction, $E_{\rm tot}=E_{\rm Dirac}^{(0)}+E_{\rm int}$. The uncertainties of the Dirac and the first-order values are determined by the nuclear size effect. It is conservatively estimated by adding quadratically two terms. The first one is obtained by varying the rms nuclear radius within its error bar. The second one estimates conservatively the uncertainty of the nuclear charge distribution by varying the distribution model from the Fermi model to the homogeneously charged sphere model. The uncertainty of the Dirac energies can be reduced if one takes into account the nuclear deformation correction, e.g., according to Ref.~\cite{Kozhedub:2008:032501}. The uncertainties of the other contributions are due to the numerical errors. They are determined by analyzing the convergence of the results with increasing the basis. The uncertainty of the total values is obtained by summing quadratically the uncertainty due to the nuclear size effect, the numerical uncertainty, and the uncertainty due to uncalculated higher-order QED contributions. In Tables~\ref{tab:total_single} and \ref{tab:total_quasi}, our theoretical predictions for the total interelectronic-interaction correction in heliumlike uranium are compared with the ones obtained in Ref.~\cite{Artemyev:2005:062104}. The results of Ref.~\cite{Artemyev:2005:062104} have been reevaluated using the new value of the rms radius~\cite{Angeli:2013:69}. One can see that there is a small discrepancy between the results which is due to the two-photon exchange and higher-order correlation contributions. 




\subsection{QED calculations of the ionization and transition energies in He-like ions \label{sec:2:2}}

\begin{figure}
\begin{center}
\includegraphics[height=4.5cm]{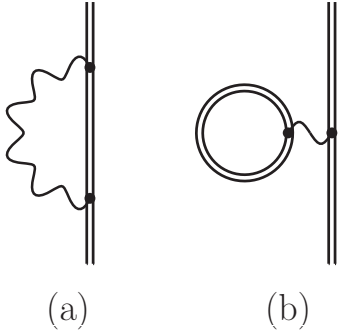}
\caption{\label{fig:se_vp}
The self-energy and vacuum-polarization diagrams.}
\end{center}
\end{figure}
\begin{figure}
\begin{center}
\includegraphics[width=13.3cm]{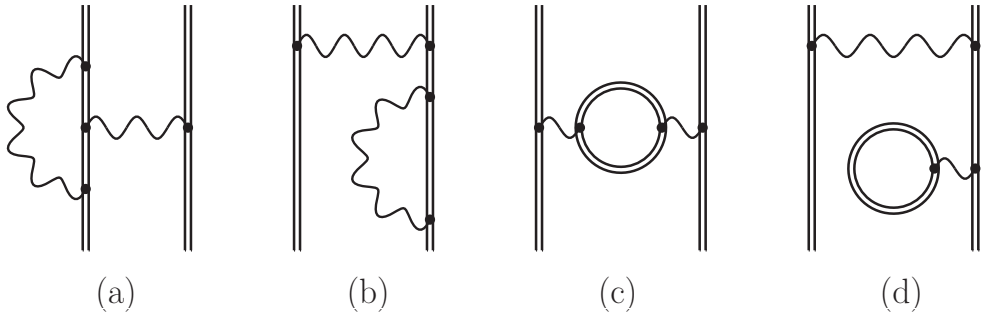}
\caption{\label{fig:scrqed}
The screened QED diagrams.}
\end{center}
\end{figure}

In Ref.~\cite{Malyshev:2019:010501_R}, we performed the QED evaluation of the four transitions from the $L$ to $K$ shell in middle-$Z$ heliumlike ions. In the present study, we extend these calculations to high-$Z$ ions. In order to complete the rigorous QED treatment in first and second orders, one has to consider the one-electron and screened QED graphs in addition to the one- and two-photon exchange Feynman diagrams shown in Figs.~\ref{fig:1ph}~and~\ref{fig:2ph_2el}. The two-loop one-electron corrections are accounted for by employing the results from Ref.~\cite{Yerokhin:2015:033103}. The contributions of the one-loop one-electron and screened QED diagrams depicted in Figs.~\ref{fig:se_vp} and \ref{fig:scrqed}, respectively, are evaluated in the present work. 
As noted in Section~\ref{sec:1:2}, the QED calculations of the energy levels can be performed with another choice of the zeroth-order approximation by including the local screening potential into the Dirac equation~(\ref{eq:dirac}). The calculations of the second-order two-electron QED contributions within this approach enable more accurate estimation of the higher-order QED corrections, which finally leads to the more precise evaluation of the energy levels in highly charged ions. In the previous section, we restricted our consideration of the correlation effects by the Coulomb potential in order to demonstrate the methods developed and compare the results obtained with the previous calculations. Nevertheless, all the methods described can readily be adopted to operate within the extended Furry picture.
As in Ref.~\cite{Malyshev:2019:010501_R}, we perform the calculations of the energy levels in high-$Z$ heliumlike ions starting from the Coulomb potential as well as by adding two different types of the screening potential into the zeroth-order approximation. We use the core-Hartree (CH) and local Dirac-Fock (LDF) screening potentials to modify the zeroth-order Hamiltonian~$h^{\rm D}$ (see, e.g., Refs.~\cite{Malyshev:2017:022512, Sapirstein:2002:042501, Shabaev:2005:062105} for the construction procedures and applications of these potentials). When applying the extended version of the Furry picture, one has to complement the diagrams shown in Figs.~\ref{fig:1ph}-\ref{fig:scrqed} with the counterterm diagrams in order to avoid double counting of the screening effects. The nuclear recoil effect on the binding energies of heliumlike ions is accounted for in the present work in accordance with the scheme described in details in Ref.~\cite{Malyshev:2018:085001}. We take into account also the nuclear polarization correction arising from the electron-nucleus interactions which include the excited intermediate nuclear states~\cite{Plunien:1991:5853, Plunien:1995:1119:note, Nefiodov:1996:227, Volotka:2014:023002, Yerokhin:2015:033103}.

The individual contributions to the binding energies of heliumlike uranium evaluated for the LDF, CH, and Coulomb potentials are shown in Table~\ref{tab:U_contrib}. As in Tables~\ref{tab:total_single} and \ref{tab:total_quasi}, the $E_{\rm Dirac}^{(0)}$ term corresponds to the energy obtained from the Dirac equation. The first- and second-order interelectronic-interaction contributions calculated employing the QED approach are given by $E_{\rm int}^{(j)}=E_{\rm Breit}^{(j)}+E_{\rm QED}^{(j)}$, where $j=1,2$. The higher-order correlation contribution evaluated within the Breit approximation is given by $E_{\rm Breit}^{(3+)}$. The one-electron and screened QED corrections are shown in columns $E^{\rm 1el}_{\rm QED}$ and $E^{\rm 2el}_{\rm ScrQED}$, respectively. The contribution of the nuclear recoil effect is presented in the column labeled with $E_{\rm rec}$. Finally, the sum of all the given contributions is shown in the last column. In Table~\ref{tab:U_contrib}, the effects of the nuclear deformation and nuclear polarization are omitted. In case of the Coulomb potential, two total values are presented. Following the scheme described in Ref.~\cite{Artemyev:2005:062104}, the second total value for the Coulomb potential is obtained by adding the higher-order QED correction $E_{\rm QED}^{\rm ho}$ evaluated according to Ref.~\cite{Drake:1988:586}. One can see that for uranium ion the inclusion of the $E_{\rm QED}^{\rm ho}$ correction increases the discrepancy between the Coulomb results and the data obtained for the screening potentials in some cases (see the related discussion in Ref.~\cite{Malyshev:2019:010501_R}). 

From Table~\ref{tab:U_contrib} it is seen that the individual terms may vary from potential to potential, whereas the total values of the binding energies obtained starting from the different initial approximation are in good agreement with each other. In this aspect, it is of interest to return once again to the discussion of the correlation effects. As noted, e.g., in Ref.~\cite{Kaygorodov:2019:032505}, the arbitrariness in the realization of the projector~$\Lambda^{(+)}$ in the DCB Hamiltonian~(\ref{eq:H_DCB}) leads to some ambiguity in the Breit-approximation results, and this ambiguity could be fully eliminated only within the rigorous QED approach. In our case, we define the projector~$\Lambda^{(+)}$ with respect to the one-electron Dirac Hamiltonian $h^{\rm D}$ which provides the zeroth-order approximation for the perturbation series. Therefore, the definition of the projector changes if we add one or the other screening potential. In the upper half of Table~\ref{tab:corr_potentials} labeled with ``DCB'', we present the binding energies of the ground and $n=2$ single states in He-like uranium evaluated within the Breit approximation for the LDF, CH, and Coulomb potentials. In the lower ``DCB$+$QED'' part, the DCB energies supplemented with the one- and two-photon exchange corrections are given. In Table~\ref{tab:U_contrib}, we omit the uncertainties of the nuclear size effect as well as the uncertainties due to the uncalculated higher-order QED corrections, and keep only the numerical ones. It is seen that the scatter of the results corresponding to the different choices of the projector~$\Lambda^{(+)}$ decreases when one takes into account the QED corrections. Moreover, the scatter of the ``DCB$+$QED'' values may serve as an alternative estimation of the uncalculated higher-order QED effects. One can conclude that the algorithm employed in Tables~\ref{tab:total_single} and \ref{tab:total_quasi} for the determination of the $E_{\rm QED}^{(3+)}$ term provides a reliable estimation.

The results of the calculations performed for the LDF screening potential have been employed as the final values of the binding energies in the present work. The energies of the mixing configurations $1s2p\, ^1P_1$ and $1s2p\, ^3P_1$ have been evaluated by diagonalizing the matrix $H$. When calculating the final results, in addition to the contributions listed in Table~\ref{tab:U_contrib}, we take into account the nuclear polarization contribution. For uranium ion, we account for the nuclear deformation correction for the Dirac energies evaluated for the Coulomb potential. In Table~\ref{tab:ionization}, we present our theoretical predictions for the ionization energies for the $n=1$ and $n=2$ states in He-like iron, xenon, and uranium. The ionization energies are obtained by subtracting the binding energies of the states under consideration from the binding energy of the $1s$ state. The results for iron are based on the calculations performed in Ref.~\cite{Malyshev:2019:010501_R} and presented here for the completeness. The theoretical uncertainties given in parentheses are estimated by summing quadratically the uncertainty due to the nuclear size effect, the numerical error of the calculations, and the uncertainty due to the uncalculated QED contributions of the higher orders. As in Ref.~\cite{Malyshev:2019:010501_R}, the uncertainty associated with the uncalculated higher-order screened QED effects is estimated in several ways. First, we study the scatter of the final results obtained for the different initial approximations. For the excited states, the maximum of the scatter for the corresponding level and the scatter for the ground state divided by the factor of 4 is taken. Second, in order to estimate the screening of the two-loop one-electron contributions by the interelectronic interaction we take the corresponding term for the $1s$ state multiplied by the factor $2/Z$. 
We have estimated the higher-order screened QED corrections by means of the model QED approach~\cite{Shabaev:2013:012513, Shabaev:2015:175, Shabaev:2018:69} as well. These estimations are found to be within the uncertainties obtained.
In Table~\ref{tab:ionization}, our theoretical predictions for the ionization energies are compared with the results obtained by Artemyev \textit{et al.} \cite{Artemyev:2005:062104}. One can see that the values presented are in agreement with each other. For high-$Z$ ions, our results have higher accuracy. On the other side, for middle-$Z$ He-like ions our uncertainties are evaluated in a more conservative way than in Ref.~\cite{Artemyev:2005:062104}.

Finally, in Table~\ref{tab:transition} we present the transition energies in He-like iron, xenon, and uranium. The transition energies were obtained as the differences of the corresponding ionization energies. Our theoretical predictions are compared with the ones obtained in Ref.~\cite{Artemyev:2005:062104}. It is seen that the values of the transition energies are in agreement with each other. For He-like uranium, our results have a higher accuracy. The theoretical predictions for the $1s2p\,^3P_2 \rightarrow 1s2s\, ^3S_1$ transition energy in He-like uranium are in good agreement with the experimental value by Trassinelli \textit{et al.}~\cite{Trassinelli:2009:63001}. As one can see from the table, the theory for high-$Z$ heliumlike ions is by one order of magnitude more accurate than the experimental value available to date.


\section{Summary}

To summarize, we have performed \textit{ab initio} QED calculations of the interelectronic-interaction corrections for the $n=1$ and $n=2$ states (including the quasi-degenerate states) in heliumlike ions. Our approach merges the rigorous QED evaluation to first and second orders in $1/Z$ with the calculations of the third- and higher-order contributions within the Breit approximation. The latter are based on the Dirac-Coulomb-Breit Hamiltonian and performed by means of two independent methods (large-scale configuration interaction and recursive perturbation theory). 
The obtained results are supplemented with the systematic estimation of the uncertainties including the contribution of the uncalculated third- and higher-order QED corrections. 
As the result, the most precise up-to-date theoretical predictions for the correlation effects in heliumlike ions are obtained. 

In addition, we have performed rigorous QED calculations of the ionization and transition energies in high-$Z$ heliumlike ion with the most advanced methods available to date. All two-electron QED corrections up to the second order are taken into account within the extended Furry picture. The nuclear recoil and nuclear polarization effects are considered as well. We have thoroughly estimated all possible sources for theoretical uncertainties. As a result, the most precise theoretical predictions for energy levels in high-$Z$ He-like ions are obtained.
In future, we plan to apply the developed approaches for QED calculations of the binding and ionization energies of the low-lying excited states in berylliumlike ions which are of current experimental interest~\cite{Bernhardt:2015:144008}.

         
\section{Acknowledgments}

This work was supported by the Russian Science Foundation (Grant No. 17-12-01097).




\begin{table}[t]
\centering

\caption{\label{tab:2ph_QED_single}
Two-photon exchange correction to the binding energies of the $n=1$ and $n=2$ single states in He-like ions, in eV.
}

\begin{tabular}{l@{\quad}
                S[table-format=-3.5(2),group-separator=]
                S[table-format=-2.5(2),group-separator=]
                S[table-format=-2.6(2),group-separator=]
                S[table-format=-2.5(2),group-separator=]
                S[table-format=-2.6(2),group-separator=]@{\quad}
                l
               }
               
\hline
\hline

  $Z$  &  {$(1s\,1s)_0$}        &  {$(1s\,2s)_0$}        &  {$(1s\,2s)_1$} 
       &  {$(1s\,2p_{1/2})_0$}  &  {$(1s\,2p_{3/2})_2$}   &  Ref.            \\
        
\hline

 18  &    -4.57813(18)  &     -3.24755(3)  &    -1.310570(4)  &     -2.13280(2)  &   -2.008834(10)  &        \\ 

   &  -4.5770  &  -3.2473  &  -1.3106  &  -2.1328  &  -2.0088  &  \cite{Artemyev:2005:062104}  \\

   &  &  -3.247532  &  -1.310570  &  &  &  \cite{Asen:2002:032516}  \\

\hline
 20  &    -4.64476(19)  &     -3.27867(3)  &    -1.315457(4)  &     -2.16820(2)  &   -2.014088(10)  &        \\ 

   &  -4.6435  &  -3.2784  &  -1.3154  &  -2.1682  &  -2.0141  &  \cite{Artemyev:2005:062104}  \\ 

   &  -4.6447  &  &             &             &             &  \cite{Blundell:1993:2615}  \\

\hline
 26  &    -4.88586(18)  &     -3.39290(4)  &    -1.333430(3)  &     -2.30013(2)  &   -2.033087(10)  &        \\ 

\hline
 30  &    -5.08122(17)  &     -3.48720(5)  &    -1.348270(3)  &     -2.41114(2)  &   -2.048399(10)  &        \\ 

   &  -5.0795  &  -3.4868  &  -1.3483  &  -2.4111  &  -2.0484  &  \cite{Artemyev:2005:062104}  \\ 

   &  -5.0812  &  &  -1.348326  &  -2.411120  &  -2.048340  &  \cite{Blundell:1993:2615,Mohr:2000:052501}  \\

   &  &  -3.487164  &  -1.348268  &  &  &  \cite{Asen:2002:032516}  \\ 

   &  &  -3.473  &  -1.348  &          &  & \cite{Andreev:2001:042513}  \\

\hline
 54  &    -6.87670(15)  &     -4.41694(6)  &    -1.492310(7)  &     -3.58493(7)  &    -2.181609(8)  &        \\ 

   &  -6.8742  &  -4.4162  &  -1.4923  &  -3.5848  &  -2.1816  &  \cite{Artemyev:2005:062104}  \\ 

\hline
 60  &    -7.51412(16)  &     -4.77222(6)  &   -1.545868(10)  &     -4.06444(8)  &    -2.225107(7)  &        \\ 

   &  -7.5114  &  -4.7714  &  -1.5459  &  -4.0642  &  -2.2251  &  \cite{Artemyev:2005:062104}  \\ 

   &  -7.5142  &  &  -1.54558   &  -4.06446   &  -2.22510   &  \cite{Blundell:1993:2615,Mohr:2000:052501}  \\

   &  &  -4.772148  &  -1.545868  &  &  &  \cite{Asen:2002:032516}  \\ 

   &  &  -4.781  &  -1.542  &  -4.068  &  & \cite{Andreev:2003:012503}  \\ 

\hline
 80  &   -10.37493(22)  &     -6.51439(9)  &   -1.796080(20)  &    -6.59554(15)  &    -2.397384(7)  &        \\ 

   & -10.3719  &  -6.5135  &  -1.7961  &  -6.5950  &  -2.3974  &  \cite{Artemyev:2005:062104}  \\ 

   &  -10.375  &  &  -1.79562   &  -6.59593   &  -2.39806   &  \cite{Blundell:1993:2615,Mohr:2000:052501}  \\

   &  &  -6.504  &  -1.789  &  -6.598  &  & \cite{Andreev:2001:042513,Andreev:2003:012503}  \\ 

\hline
 92  &   -12.87444(45)  &    -8.21322(22)  &   -2.022063(25)  &    -9.27103(21)  &    -2.522774(8)  &        \\ 

   & -12.8714  &  -8.2122  &  -2.0221  &  -9.2701  &  -2.5228  &  \cite{Artemyev:2005:062104}  \\ 

   &           &  &  -2.02034   &  -9.27598   &  -2.52228   &  \cite{Mohr:2000:052501}  \\

   &  &  -8.213058  &  -2.021988  &  &  &  \cite{Asen:2002:032516}  \\ 

   &  &  -8.184  &  -2.018  &  -9.274  &  & \cite{Andreev:2001:042513,Andreev:2003:012503}  \\ 

\hline

\hline
\hline

\end{tabular}%

\end{table}

\begin{table}[t]
\centering

\caption{\label{tab:2ph_QED_quasi}
Two-photon exchange contributions to the matrix $H_{ik}$ for the $n=2$ quasi-degenerate states in He-like ions, in eV. 
$(1s\,2p_{1/2})_1$ and $(1s\,2p_{3/2})_1$ stand for the corresponding diagonal matrix elements of the operator $H$, whereas ``off-diag.'' refers to the off-diagonal matrix elements
evaluated according to Eqs.~(\ref{eq:H2_ik}) and (\ref{eq:H2_ik_total}). See the text for details.
}

\begin{tabular}{l@{\quad}
                S[table-format=-2.5(2),group-separator=]
                S[table-format=-2.5(2),group-separator=]
                S[table-format=-2.5(2),group-separator=]
                S[table-format=-2.5(2),group-separator=]@{\quad}
                l
               }
               
\hline
\hline

  $Z$  &  {$(1s\,2p_{1/2})_1$}   &  {$(1s\,2p_{3/2})_1$}   &  {off-diag.} 
       &  {off-diag.}           &  Ref.            \\
        &                        &                        &  {Eq.~(\ref{eq:H2_ik})}
       &  {Eq.~(\ref{eq:H2_ik_total})}   &     \\
        
\hline

 18  &     -2.81743(2)  &     -3.56150(4)  &     -1.06268(6)  &     -1.06268(6)  &        \\ 

   &  -2.8173  &  -3.5613  &  -1.0626  &    &  \cite{Artemyev:2005:062104}  \\

   &  -2.8168  &  -3.5603  &    &  -1.0618  & \cite{Andreev:2004:062505}  \\

\hline
 20  &     -2.83385(2)  &     -3.57346(5)  &     -1.05907(5)  &     -1.05908(5)  &        \\ 

   &  -2.8337  &  -3.5733  &  -1.0589  &    &  \cite{Artemyev:2005:062104}  \\

\hline
 26  &     -2.89433(4)  &     -3.61710(8)  &     -1.04592(4)  &     -1.04595(4)  &        \\ 

   &  -2.8938  &  -3.6142  &    &  -1.0450  & \cite{Andreev:2004:062505}  \\ 

\hline
 30  &     -2.94445(5)  &    -3.65280(10)  &     -1.03521(3)  &     -1.03527(3)  &        \\ 

   &  -2.9443  &  -3.6525  &  -1.0350  &    &  \cite{Artemyev:2005:062104}  \\

   &  -2.9439  &  -3.6506  &    &  -1.0350  & \cite{Andreev:2004:062505}  \\ 

\hline
 54  &     -3.44310(7)  &    -3.98749(17)  &     -0.93906(6)  &     -0.93966(6)  &        \\ 

   &  -3.4429  &  -3.9871  &  -0.9387  &    &  \cite{Artemyev:2005:062104}  \\

\hline
 60  &     -3.63509(8)  &    -4.10707(17)  &     -0.90671(6)  &     -0.90760(6)  &        \\ 

   &  -3.6348  &  -4.1066  &  -0.9064  &    &  \cite{Artemyev:2005:062104}  \\

   &  -3.635   &  -4.105   &    &  -0.893   & \cite{Andreev:2004:062505}  \\ 

\hline
 80  &    -4.58691(13)  &    -4.63156(17)  &    -0.77545(11)  &    -0.77781(11)  &        \\ 

   &  -4.5866  &  -4.6312  &  -0.7752  &    &  \cite{Artemyev:2005:062104}  \\

   &  -4.585   &  -4.628   &    &  -0.771   & \cite{Andreev:2004:062505}  \\  

\hline
 92  &    -5.53336(21)  &    -5.05529(18)  &    -0.67877(12)  &    -0.68190(12)  &        \\ 

   &  -5.5329  &  -5.0550  &  -0.6787  &    &  \cite{Artemyev:2005:062104}  \\

   &  -5.531   &  -5.053   &    &  -0.683   & \cite{Andreev:2004:062505}  \\ 

\hline

\hline
\hline

\end{tabular}%

\end{table}


\clearpage

\begin{table}[t]
\centering

\caption{\label{tab:total_single}
Zeroth-order contribution and interelectronic-interaction corrections to the binding energies of the $n=1$ and $n=2$ single states in He-like ions, in eV.
}

\resizebox{\textwidth}{!}{%
\begin{tabular}{l@{\quad}
                l
                S[table-format=-6.5(2),group-separator=,table-align-text-post=false]
                S[table-format=-6.5(2),group-separator=,table-align-text-post=false]
                S[table-format=-6.6(2),group-separator=,table-align-text-post=false]
                S[table-format=-6.5(2),group-separator=,table-align-text-post=false]
                S[table-format=-6.5(2),group-separator=,table-align-text-post=false]
               }
               
\hline
\hline

  Nucl.  &  Contrib.  &  \multicolumn{1}{c}{$(1s\,1s)_0$}              &  
                         \multicolumn{1}{c}{$(1s\,2s)_0$}        &  
                         \multicolumn{1}{c}{$(1s\,2s)_1$}        &  
                         \multicolumn{1}{c}{$(1s\,2p_{1/2})_0$}   &  
                         \multicolumn{1}{c}{$(1s\,2p_{3/2})_2$}    \\
        
\hline

  $^{56}_{26}{\rm Fe}$ & $E_{\rm Dirac}^{(0)}$  &       -18563.38437(11) &        -11607.41420(6) &        -11607.41420(6) &        -11607.42096(5) &        -11586.25098(5)     \\ 
                       & $E_{\rm Breit}^{(1)}$  &           454.61698(1) &              168.42821 &             134.957868 &              165.43432 &             160.747756     \\ 
                       & $E_{\rm QED}^{(1)}$    &                    0.0 &                0.00140 &               0.000468 &               -0.00123 &              -0.013014     \\ 
                       & $E_{\rm Breit}^{(2)}$  &            -4.89911(9) &            -3.39621(1) &           -1.333235(3) &            -2.30222(1) &           -2.034093(9)     \\ 
                       & $E_{\rm QED}^{(2)}$    &             0.01325(9) &             0.00331(3) &           -0.000196(1) &                0.00209 &            0.001007(1)     \\ 
                       & $E_{\rm Breit}^{(3+)}$ &             0.02516(4) &             0.01727(3) &           -0.004815(1) &               -0.01340 &           -0.017656(1)     \\ 
                       & $E_{\rm QED}^{(3+)}$   &       \pm      0.00014 &       \pm      0.00014 &       \pm      0.00014 &       \pm      0.00014 &       \pm      0.00014     \\ 
                       & $E_{\rm int}$          &          449.75628(26) &          165.05398(15) &          133.62009(14) &          163.11957(14) &          158.68400(14)     \\ 
                       & $E_{\rm tot}$          &       -18113.62809(28) &       -11442.36022(16) &       -11473.79411(15) &       -11444.30138(15) &       -11427.56698(15)     \\ 

\hline

 $^{132}_{54}{\rm Xe}$ & $E_{\rm Dirac}^{(0)}$  &         -82687.590(14) &        -51786.8006(83) &        -51786.8006(83) &        -51787.2428(72) &        -51360.5206(72)     \\ 
                       & $E_{\rm Breit}^{(1)}$  &         1036.55776(24) &           383.43652(6) &         295.372824(34) &           387.58771(2) &         341.414191(11)     \\ 
                       & $E_{\rm QED}^{(1)}$    &                    0.0 &                0.05252 &               0.017508 &               -0.03052 &              -0.503357     \\ 
                       & $E_{\rm Breit}^{(2)}$  &            -7.04127(7) &            -4.46021(2) &           -1.495551(1) &            -3.63599(1) &           -2.210487(4)     \\ 
                       & $E_{\rm QED}^{(2)}$    &             0.16457(8) &             0.04327(4) &            0.003241(5) &             0.05106(6) &            0.028878(4)     \\ 
                       & $E_{\rm Breit}^{(3+)}$ &             0.04198(8) &             0.02362(4) &           -0.001328(1) &             0.00819(1) &           -0.007960(1)     \\ 
                       & $E_{\rm QED}^{(3+)}$   &       \pm       0.0020 &       \pm       0.0020 &       \pm       0.0020 &       \pm       0.0020 &       \pm       0.0020     \\ 
                       & $E_{\rm int}$          &          1029.7230(20) &           379.0957(20) &           293.8967(20) &           383.9805(20) &           338.7213(20)     \\ 
                       & $E_{\rm tot}$          &         -81657.867(15) &        -51407.7049(85) &        -51492.9039(85) &        -51403.2624(75) &        -51021.7993(75)     \\ 

\hline

  $^{238}_{92}{\rm U}$ & $E_{\rm Dirac}^{(0)}$  &         -264162.56(83) &         -166259.03(50) &         -166259.03(50) &         -166292.35(43) &         -161731.12(42)     \\ 
                       & $E_{\rm Breit}^{(1)}$  &           2265.887(12) &           849.4612(40) &           587.9448(17) &           922.8337(14) &          618.12333(36)     \\ 
                       & $E_{\rm QED}^{(1)}$    &                    0.0 &                0.67547 &            0.225155(1) &                0.36470 &          -7.196269(20)     \\ 
                       & $E_{\rm Breit}^{(2)}$  &          -14.15639(13) &            -8.57043(4) &           -2.074590(8) &            -9.75205(4) &           -2.790592(3)     \\ 
                       & $E_{\rm QED}^{(2)}$    &            1.28195(32) &            0.35722(18) &           0.052527(17) &            0.48102(16) &            0.267818(5)     \\ 
                       & $E_{\rm Breit}^{(3+)}$ &            0.09633(15) &             0.05801(5) &            0.001969(2) &             0.07502(2) &           -0.002922(1)     \\ 
                       & $E_{\rm QED}^{(3+)}$   &       \pm        0.017 &       \pm        0.017 &       \pm        0.017 &       \pm        0.017 &       \pm        0.017     \\ 
                       & $E_{\rm int}$          &           2253.109(21) &            841.981(18) &            586.150(18) &            914.002(18) &            608.401(17)     \\ 
                       &                        &    2253.079$^{\,\ddagger}$ &     841.953$^{\,\ddagger}$ &     586.148$^{\,\ddagger}$ &     913.935$^{\,\ddagger}$ &     608.399$^{\,\ddagger}$     \\ 
                       & $E_{\rm tot}$          &         -261909.46(83) &         -165417.05(50) &         -165672.88(50) &         -165378.35(43) &         -161122.72(42)     \\ 

\hline

\hline
\hline

\end{tabular}%
}

{
\begin{flushleft}
$^{\ddagger}$ Artemyev \textit{et al.}~\cite{Artemyev:2005:062104}. The results are corrected for the updated value of the root-mean-square nuclear radius from Ref.~\cite{Angeli:2013:69}.
\end{flushleft}
}

\end{table}

\begin{table}[t]
\centering

\renewcommand{\arraystretch}{0.9}

\caption{\label{tab:total_quasi}
Zeroth-order contribution and interelectronic-interaction corrections to the matrix $H_{ik}$ for the $n=2$ quasi-degenerate states in He-like ions, in eV.
$(1s\,2p_{1/2})_1$ and $(1s\,2p_{3/2})_1$ stand for the corresponding diagonal matrix elements of the operator $H$, whereas off-diag.'' refers to the off-diagonal matrix elements.
}

\begin{tabular}{l@{\quad}
                l
                S[table-format=-6.5(2),group-separator=,table-align-text-post=false]
                S[table-format=-6.5(2),group-separator=,table-align-text-post=false]
                S[table-format=-3.5(2),group-separator=,table-align-text-post=false]
               }
               
\hline
\hline

  Nucl.  &  Contrib.  &  \multicolumn{1}{c}{$(1s\,2p_{1/2})_1$}   &  
                         \multicolumn{1}{c}{$(1s\,2p_{3/2})_1$}   &  
                         \multicolumn{1}{c}{off-diag.}           \\
        
\hline

  $^{56}_{26}{\rm Fe}$ & $E_{\rm Dirac}^{(0)}$  &        -11607.42096(5) &        -11586.25098(5) &                    0.0     \\ 
                       & $E_{\rm Breit}^{(1)}$  &              171.21174 &              177.16373 &               10.72632     \\ 
                       & $E_{\rm QED}^{(1)}$    &               -0.00041 &                0.00524 &                0.00799     \\ 
                       & $E_{\rm Breit}^{(2)}$  &            -2.89462(1) &            -3.61679(3) &            -1.04482(1)     \\ 
                       & $E_{\rm QED}^{(2)}$    &             0.00029(4) &            -0.00031(5) &            -0.00113(3)     \\ 
                       & $E_{\rm Breit}^{(3+)}$ &            -0.00004(1) &             0.01696(2) &             0.02294(1)     \\ 
                       & $E_{\rm QED}^{(3+)}$   &       \pm      0.00014 &       \pm      0.00014 &       \pm      0.00014     \\ 
                       & $E_{\rm int}$          &          168.31696(15) &          173.56883(17) &            9.71130(14)     \\ 
                       & $E_{\rm tot}$          &       -11439.10400(16) &       -11412.68215(18) &            9.71130(14)     \\ 

\hline

 $^{132}_{54}{\rm Xe}$ & $E_{\rm Dirac}^{(0)}$  &        -51787.2428(72) &        -51360.5206(72) &                    0.0     \\ 
                       & $E_{\rm Breit}^{(1)}$  &           382.11692(1) &              377.42038 &            17.59572(1)     \\ 
                       & $E_{\rm QED}^{(1)}$    &               -0.01017 &                0.20549 &                0.30647     \\ 
                       & $E_{\rm Breit}^{(2)}$  &            -3.45773(2) &            -3.98282(7) &            -0.91310(3)     \\ 
                       & $E_{\rm QED}^{(2)}$    &             0.01463(5) &            -0.00468(9) &            -0.02656(3)     \\ 
                       & $E_{\rm Breit}^{(3+)}$ &             0.00515(1) &             0.01606(3) &             0.01293(2)     \\ 
                       & $E_{\rm QED}^{(3+)}$   &       \pm       0.0020 &       \pm       0.0020 &       \pm       0.0020     \\ 
                       & $E_{\rm int}$          &           378.6688(20) &           373.6544(20) &            16.9755(20)     \\ 
                       & $E_{\rm tot}$          &        -51408.5740(75) &        -50986.8661(75) &            16.9755(20)     \\ 

\hline

  $^{238}_{92}{\rm U}$ & $E_{\rm Dirac}^{(0)}$  &         -166292.35(43) &         -161731.12(42) &                    0.0     \\ 
                       & $E_{\rm Breit}^{(1)}$  &          809.58318(90) &           683.97647(2) &            8.66822(25)     \\ 
                       & $E_{\rm QED}^{(1)}$    &                0.12157 &             3.02676(1) &             4.26371(2)     \\ 
                       & $E_{\rm Breit}^{(2)}$  &            -5.71256(5) &            -4.98203(9) &            -0.45327(6)     \\ 
                       & $E_{\rm QED}^{(2)}$    &            0.17920(16) &            -0.07326(9) &            -0.22863(6)     \\ 
                       & $E_{\rm Breit}^{(3+)}$ &             0.01722(1) &             0.02370(2) &             0.00848(1)     \\ 
                       & $E_{\rm QED}^{(3+)}$   &       \pm        0.017 &       \pm        0.017 &       \pm        0.017     \\ 
                       & $E_{\rm int}$          &            804.189(17) &            681.972(17) &             12.259(17)     \\ 
                       &                        &     804.180$^{\,\ddagger}$ &     681.969$^{\,\ddagger}$ &      12.267$^{\,\ddagger}$     \\ 
                       & $E_{\rm tot}$          &         -165488.16(43) &         -161049.15(42) &             12.259(17)     \\ 

\hline

\hline
\hline

\end{tabular}%

{
\begin{flushleft}
$^{\ddagger}$ Artemyev \textit{et al.}~\cite{Artemyev:2005:062104}. The results are corrected for the updated value of the root-mean-square nuclear radius from Ref.~\cite{Angeli:2013:69}.
\end{flushleft}
}

\end{table}

\begin{table}[t]
\centering

\renewcommand{\arraystretch}{0.83}

\caption{\label{tab:U_contrib}
Individual contributions to the binding energies of He-like uranium evaluated for the local Dirac-Fock (LDF), core-Hartree (CH), and Coulomb (Coul) potentials as the initial approximations, in eV. The nuclear deformation and nuclear polarization corrections are omitted. For the quasi-degenerate states, the contributions to the $H$-matrix elements are listed. 
}

\resizebox{\textwidth}{!}{%
\begin{tabular}{l@{\quad}
                l@{\quad}
                S[table-format=-6.4]
                S[table-format=-4.4]
                S[table-format=-2.4]
                S[table-format=-1.5]
                S[table-format= 3.4]
                S[table-format=-1.4]
                S[table-format= 1.4]
                S[table-format=-6.4, table-align-text-post=false]
                }

\hline
\hline

 State  &  $V_{\rm eff}$
        &  {$ E_{\rm Dirac}^{(0)}$}  &  {$ E^{(1)}_{\rm int}$}  &  {$ E^{(2)}_{\rm int}$}  &
                    {$ E^{(3+)}_{\rm Breit}$}  &  {$ E^{\rm 1el}_{\rm QED}$}  &  
                    {$ E^{\rm 2el}_{\rm ScrQED}$}  &  {$ E_{\rm rec}$}  &  {Sum} \\

\hline

 $(1s\,1s)_0$       & LDF  &  -260754.077  &  -1150.188  &  -5.264  &  0.060  &  525.922  &  -2.915  &  0.931 &  -261385.531             \\
                  & CH   &  -260311.808  &  -1593.468  &  -4.245  &  0.047  &  524.714  &  -1.697  &  0.930 &  -261385.526             \\ 
                  & Coul &  -264162.564  &   2265.887  & -12.874  &  0.096  &  530.131  &  -7.151  &  0.935 &  -261385.540             \\ 
                  &      &               &             &          &         &           &          &        &  -261385.591$^\ddagger$  \\  
                                                                                                              
\hline                                                                                                              
                                                                                                              
 $(1s\,2s)_0$       & LDF  &  -163906.247  &  -1507.099  &  -3.739  &  0.029  &  311.555  &   0.903  &  0.585 &  -165104.014             \\ 
                  & CH   &  -163655.001  &  -1757.754  &  -4.332  &  0.030  &  310.858  &   1.602  &  0.584 &  -165104.013             \\ 
                  & Coul &  -166259.030  &    850.137  &  -8.213  &  0.058  &  314.601  &  -2.168  &  0.588 &  -165104.028             \\ 
                  &      &               &             &          &         &           &          &        &  -165104.038$^\ddagger$  \\  
                                                                                                              
\hline                                                                                                               
                                                                                                              
 $(1s\,2s)_1$       & LDF  &  -163906.247  &  -1765.402  &  -1.237  &  0.005  &  311.555  &   2.030  &  0.586 &  -165358.711             \\ 
                  & CH   &  -163655.001  &  -2015.782  &  -2.107  &  0.008  &  310.858  &   2.728  &  0.585 &  -165358.711             \\ 
                  & Coul &  -166259.030  &    588.170  &  -2.022  &  0.002  &  314.601  &  -1.023  &  0.589 &  -165358.714             \\ 
                  &      &               &             &          &         &           &          &        &  -165358.711$^\ddagger$  \\  
                                                                                                              
 \hline                                                                                                              
                                                                                                              
 $(1s\,2p_{1/2})_0$ & LDF  &  -163805.880  &  -1570.384  &  -2.117  &  0.015  &  269.491  &   1.534  &  0.521 &  -165106.821             \\ 
                  & CH   &  -163550.249  &  -1825.675  &  -2.457  &  0.011  &  268.846  &   2.182  &  0.520 &  -165106.821             \\ 
                  & Coul &  -166292.353  &    923.198  &  -9.271  &  0.075  &  271.912  &  -0.911  &  0.523 &  -165106.827             \\ 
                  &      &               &             &          &         &           &          &        &  -165106.826$^\ddagger$  \\  
                                                                                                              
 \hline                                                                                                              
                                                                                                              
 $(1s\,2p_{3/2})_2$ & LDF  &  -159398.331  &  -1723.089  &  -1.305  &  0.004  &  271.414  &   1.918  &  0.500 &  -160848.888             \\ 
                  & CH   &  -159161.309  &  -1959.201  &  -2.219  &  0.008  &  270.780  &   2.553  &  0.499 &  -160848.889             \\ 
                  & Coul &  -161731.119  &    610.927  &  -2.523  & -0.003  &  273.841  &  -0.517  &  0.502 &  -160848.892             \\ 
                  &      &               &             &          &         &           &          &        &  -160848.885$^\ddagger$  \\ 
                  
 \hline                                                                                                              
                                                                                                              
 $(1s\,2p_{1/2})_1$ & LDF  &  -163805.880  &  -1680.798  &  -1.501  &  0.008  &  269.491  &   1.769  &  0.541 &  -165216.369             \\ 
                  & CH   &  -163550.249  &  -1935.653  &  -2.281  &  0.011  &  268.846  &   2.416  &  0.541 &  -165216.370             \\ 
                  & Coul &  -166292.353  &    809.705  &  -5.533  &  0.017  &  271.912  &  -0.668  &  0.544 &  -165216.377             \\ 
                  &      &               &             &          &         &           &          &        &  -165216.376$^\ddagger$  \\  
                                                                                                              
 \hline                                                                                                              
                                                                                                              
 $(1s\,2p_{3/2})_1$ & LDF  &  -159398.331  &  -1648.768  &  -2.062  &  0.013  &  271.414  &   1.995  &  0.545 &  -160775.192             \\ 
                  & CH   &  -159161.309  &  -1884.975  &  -2.879  &  0.017  &  270.780  &   2.630  &  0.545 &  -160775.192             \\ 
                  & Coul &  -161731.119  &    687.003  &  -5.055  &  0.024  &  273.841  &  -0.441  &  0.547 &  -160775.200             \\ 
                  &      &               &             &          &         &           &          &        &  -160775.202$^\ddagger$  \\   
                                                                                                              
 \hline                                                                                                              
                                                                                                              
 off-diag.        & LDF  &        0.000  &     12.759  &  -0.497  &  0.006  &    0.000  &   0.094  &  0.030 &       12.392             \\ 
                  & CH   &        0.000  &     12.781  &  -0.519  &  0.006  &    0.000  &   0.094  &  0.030 &       12.393             \\ 
                  & Coul &        0.000  &     12.932  &  -0.682  &  0.008  &    0.000  &   0.094  &  0.030 &       12.383             \\
                  &      &               &             &          &         &           &          &        &       12.378$^\ddagger$  \\    

\hline
\hline

\end{tabular}%
}

{
\begin{flushleft}
$^\ddagger$ The higher-order QED correction $\Delta E^{\rm ho}_{\rm QED}$ evaluated according to Refs.~\cite{Drake:1988:586, Artemyev:2005:062104} is added to the total Coulomb value, see the details in Ref.~\cite{Malyshev:2019:010501_R}.
\end{flushleft}
}

\end{table}

\begin{table}[t]
\centering

\renewcommand{\arraystretch}{0.95}

\caption{\label{tab:corr_potentials}
Interelectronic-interaction contributions to the binding energies of the $n=1$ and $n=2$ single states in He-like uranium evaluated for the local Dirac-Fock (LDF), core-Hartree (CH), and Coulomb (Coul) potentials, in eV. The ``DCB'' stands for the calculations based on the Dirac-Coulomb-Breit Hamiltonian~(\ref{eq:DCB}) and its generalization to the case of the extended Furry picture. The ``DCB$+$QED''-values include the one- and two-photon exchange QED contributions. All the given digits are correct within the numerical uncertainty. The uncertainty of the nuclear size effect and the uncertainty due to uncalculated higher-order QED contributions are omitted.}

\begin{tabular}{c@{\quad}
                l
                S[table-format=-8.3]
                S[table-format=-7.4]
                S[table-format=-7.4]
                S[table-format=-7.4]
                S[table-format=-7.3]
                }
                
\hline
\hline

 Contrib.   &  {$V_{\rm eff}$}     &  {$(1s\,1s)_0$}  &  {$(1s\,2s)_0$}  &  {$(1s\,2s)_1$}  
            &  {$(1s\,2p_{1/2})_0$}  &  {$(1s\,2p_{3/2})_2$}  \\
            
\hline

           &  LDF  & -261910.814 & -165418.107  & -165673.170  & -165379.129  & -161115.807  \\
 DCB       &  CH   & -261910.854 & -165418.121  & -165673.183  & -165379.123  & -161115.821  \\
           &  Coul & -261910.737 & -165418.081  & -165673.158  & -165379.197  & -161115.789  \\
   
\hline   
   
           &  LDF  & -261909.469 & -165417.056  & -165672.882  & -165378.367  & -161122.721  \\
 DCB$+$QED &  CH   & -261909.473 & -165417.057  & -165672.882  & -165378.369  & -161122.721  \\
           &  Coul & -261909.455 & -165417.049  & -165672.880  & -165378.351  & -161122.717  \\
 
\hline
\hline

\end{tabular}%

\end{table}

\begin{table}[t]
\centering

\renewcommand{\arraystretch}{0.95}

\caption{\label{tab:ionization}
Ionization energies for the $n=1$ and $n=2$ states in He-like ions, in eV, and comparison with Artemyev \textit{et al.}~\cite{Artemyev:2005:062104}. The results of Ref.~\cite{Artemyev:2005:062104} are reevaluated using the CODATA 2014 recommended values of the fundamental constants~\cite{Mohr:2016:035009} and corrected for the updated value of the uranium root-mean-square nuclear radius from Ref.~\cite{Angeli:2013:69}.}

\resizebox{\textwidth}{!}{%
\begin{tabular}{@{}l
                S[table-format=7.4(2)]
                S[table-format=6.4(2)]
                S[table-format=6.4(2)]
                S[table-format=6.4(2)]
                S[table-format=6.4(2)]
                S[table-format=6.4(2)]
                S[table-format=6.5(2)]
                l
                @{}}
                
\hline
\hline

  $Z$    &    {$1s^2\,^1S_0$}     &    {$1s2s\,^1S_0$}     &    {$1s2s\,^3S_1$}     &
              {$1s2p\,^3P_0$}     &    {$1s2p\,^3P_1$}     &    {$1s2p\,^1P_1$}     &    {$1s2p\,^3P_2$}    &    Ref.   \\

\hline

26  &  8828.1896(25)  &   2160.1625(8)  &   2191.5742(7)  &   2162.6253(7)  &   2160.6082(7)  &   2127.7523(9)  &   2145.8530(7)  &                               \\

    &  8828.1875(11)  &   2160.1632(7)  &   2191.5745(6)  &  2162.6261(10)  &   2160.6085(4)  &   2127.7524(2)  &   2145.8532(2)  &  \cite{Artemyev:2005:062104}  \\
       
\hline       
       
54  &  40271.726(26)  & 10057.5425(78)  & 10142.5688(70)  & 10059.4870(65)  & 10065.4536(65)  &  9641.6669(65)  &  9677.3570(65)  &                               \\

    &  40271.722(16)  &  10057.561(32)  &  10142.579(32)  &  10059.496(12)  & 10065.4554(80)  &  9641.6693(72)  &  9677.3570(72)  &  \cite{Artemyev:2005:062104}  \\
       
\hline  
       
92  &  129570.09(53)  &   33288.33(10)  &   33543.02(10)  &  33291.081(43)  &  33400.664(43)  &  28959.411(39)  &  29033.142(39)  &                               \\

    &  129570.62(64)  &   33288.57(24)  &   33543.21(24)  &   33291.15(13)  &   33400.68(11)  &   28959.41(10)  &   29033.12(10)  &  \cite{Artemyev:2005:062104}  \\

\hline
\hline

\end{tabular}%
}

\end{table}

\begin{table}[t]
\centering

\renewcommand{\arraystretch}{0.95}

\caption{\label{tab:transition}
Transition energies in He-like ions, in eV, and comparison with Artemyev \textit{et al.}~\cite{Artemyev:2005:062104}. The results of Ref.~\cite{Artemyev:2005:062104} are reevaluated using the CODATA 2014 recommended values of the fundamental constants~\cite{Mohr:2016:035009} and corrected for the updated value of the uranium root-mean-square nuclear radius from Ref.~\cite{Angeli:2013:69}.}

\begin{tabular}{l
                S[table-format=7.4(2)]
                S[table-format=6.4(2)]
                S[table-format=6.4(2)]
                S[table-format=6.4(2)]
                S[table-format=6.4(2)]@{\quad}
                l
                }
                
\hline
\hline


  $Z$    &    {$1s2p\,^1P_1 $}     &    {$1s2p\,^3P_2 $}     &    
              {$1s2p\,^3P_1 $}     &    {$1s2s\,^3S_1 $}     &    
              {$1s2p\,^3P_2 $}     &    Ref.   \\
         &    {$ \rightarrow 1s^2\,^1S_0$}     &    {$ \rightarrow 1s^2\,^1S_0$}     &    
              {$ \rightarrow 1s^2\,^1S_0$}     &    {$ \rightarrow 1s^2\,^1S_0$}     &    
              {$ \rightarrow 1s2s\,^3S_1$}     &       \\              

\hline

26  &   6700.4373(25)  &   6682.3366(25)  &   6667.5814(25)  &   6636.6154(25)  &   45.7212(10)   &                                \\

    &   6700.4351(11)  &   6682.3343(11)  &   6667.5790(12)  &   6636.6130(13)  &    45.7213(6)   &  \cite{Artemyev:2005:062104}   \\
                                                                                 
\hline                                                                                 
                                                                                 
54  &   30630.059(27)  &   30594.369(27)  &   30206.273(27)  &   30129.157(27)  &  465.2119(96)   &                                \\ 

    &  30630.053(18)   &  30594.365(18)   &  30206.267(18)   &  30129.143(36)   &  465.222(33)    &  \cite{Artemyev:2005:062104}   \\
                                                                                 
\hline   
   
92  &   100610.68(54)  &   100536.95(54)  &    96169.43(54)  &    96027.07(54)  &   4509.88(11)   &                                \\ 
 
    &   100611.21(65)  &   100537.50(65)  &    96169.94(65)  &    96027.41(68)  &   4510.09(26)   &  \cite{Artemyev:2005:062104}   \\
         
    &                  &                  &                  &                  &   4509.71(99)   &  \cite{Trassinelli:2009:63001} \\
     
\hline
\hline

\end{tabular}%

\end{table}

\end{document}